\documentclass[aps,10pt,prc,superscriptaddress, nofootinbib, twocolumn,tightenlines]{revtex4-1}
\usepackage{graphicx,subfigure}
\usepackage{amsmath,amssymb}
\usepackage{multirow}
\usepackage{bm}
\usepackage{verbatim}

\newcommand{\be}{\begin{eqnarray}}
\newcommand{\ee}{\end{eqnarray}}
\newcommand{\nn}{\nonumber}

\def\slashchar#1{\setbox0=\hbox{$#1$}           % set a box for #1 
   \dimen0=\wd0                                 % and get its size
   \setbox1=\hbox{/} \dimen1=\wd1               % get size of /
  \ifdim\dimen0>\dimen1                        % #1 is bigger
 \rlap{\hbox to \dimen0{\hfil/\hfil}}      % so center / in box
  #1                                        % and print #1
 \else                                        % / is bigger
    \rlap{\hbox to \dimen1{\hfil$#1$\hfil}}   % so center #1
    /                                         % and print /
 \fi}                                         %

\begin{document}

\title{Baryon preclustering at the freeze-out of heavy-ion collisions \\ and light-nuclei production}

\author{Edward Shuryak} 
\affiliation{Department of Physics and Astronomy, Stony Brook University,
Stony Brook, NY 11794-3800, USA}
\author{Juan M. Torres-Rincon}
\affiliation{Institut f\"ur Theoretische Physik, Johann Wolfgang Goethe-Universit\"at, Max-von-Laue-Strasse 1, D-60438 Frankfurt am Main, Germany}
\affiliation{Department of Physics and Astronomy, Stony Brook University,
Stony Brook, NY 11794-3800, USA}

\begin{abstract}
Following the idea of nucleon clustering and light-nuclei production in relativistic heavy-ion collisions close to the QCD critical-end point, we address the quantum effects affecting the interaction of several nucleons at finite temperature. For this aim we use the $K$-harmonics method to four-nucleon states ($\alpha$ particle), and also develop a novel semiclassical  ``flucton'' method at finite temperature, based on certain classical paths in Euclidean time, and apply it to two- and four-particle configurations. To study possible effects on the light-nuclei production close to the QCD critical point, we also made such calculations with modified internuclear potentials. For heavy-ion experiments, we propose new measurements of light-nuclei multiplicity ratios which may show enhancements due to baryon preclustering. We point out the special role of the $\mathcal{O}(50)$ four-nucleon excitations of $\alpha$-particle, feeding into the final multiplicities of $d,t$, $^3$He, and $^4$He, and propose to directly look for their 2-body decays.
\end{abstract}

\maketitle

\section{Introduction}

In the past decade the physics of heavy-ion collisions has significantly widened its scope. Traditional studies of Au+Au and Pb+Pb collisions at the highest energies of the RHIC (Relativistic Heavy-Ion Collider) and the LHC (Large Hadron Collider) have quantified the unusual properties of the quark-gluon plasma. A significant progress was reached in studies of ``small systems'', central $p+A$ and high-multiplicity $p+p$, in which radial, elliptic and triangular flows have been observed, confirming hydrodynamical explosions at sufficiently large multiplicities~\cite{Nagle:2018nvi}. The final particle composition is well described close to the phase transition line by the so-called statistical hadronization models~\cite{Andronic:2017pug}.

With these progresses at the high-energy frontier, there is a growing interest in better understanding the lower collision energies, related to larger baryonic densities. The suggestion of the possible existence of the QCD critical point and therefore increased event-by-event fluctuations~\cite{Stephanov:1998dy} has lead to the RHIC Beam-Energy Scan (BES) program. Complementing it, dedicated low-energy colliders are under construction in Germany (FAIR-GSI, Darmstadt) and Russia (NICA, Dubna), with similar projects under consideration in China (HIAF) and Japan (J-PARC-HI).

At this time,  experiments show two particular intriguing observations which might be related to the QCD critical point. One is the significant modification of the shape of net-proton multiplicity distribution (large scaled kurtosis) at the lowest RHIC collision energies observed by STAR collaboration~\cite{Luo:2015ewa}. Another (to be shown in Fig.~\ref{fig_ratio_tpdd}) is an apparent increase at $\sqrt{s_{NN}}=20-30$ GeV of the tritium production relative to deuterium and to the statistical hadronization model in the same experiment~\cite{Liu:2019ppd}. 

In our previous paper~\cite{Shuryak:2018lgd} we put forward the idea that a sizable scaled 
kurtosis of the proton distribution is another aspect of the preclustering of nucleons (or prenuclei) at the freeze-out stage, due to the modification of $NN$ potential. This effect would also lead to an increase of light-nuclei production with respect to the statistical model expectations. Let us present some qualitative arguments emphasizing the main points made in Ref.~\cite{Shuryak:2018lgd}.

To begin with let us compare the situation at the freeze-out of high-energy heavy-ion collisions with other known situations in which various nuclear fragments---and especially light nuclei---are known to be produced~\cite{Braun-Munzinger:2018hat,Vovchenko:2019aoz}. In particular, their natural production is known to occur in the Big Bang, and later in stars.  In these cases the temperature $T$ is much lower than the binding energy of the states, $T\ll B$\footnote{We will use natural units in this paper, $\hbar=c=k_B=1$. In some places of this paper we will make $\hbar$ explicit.}, and the corresponding Boltzmann factors $\exp( B/T )$ are large and play a crucial role. In heavy-ion collisions at semirelativistic energies, one has $T\sim B$ and the production of nuclear fragments.   

Unlike these conditions, the freeze-out temperatures we will consider are large compared to the binding energy 
\be B\sim \textrm{ few MeV } \ll T=100 - 150 \textrm{ MeV} \nn \ , \ee
and therefore the binding energies of the resulting light nuclei are basically irrelevant. 

\begin{figure}[ht]
\begin{center}
\includegraphics[width=7cm]{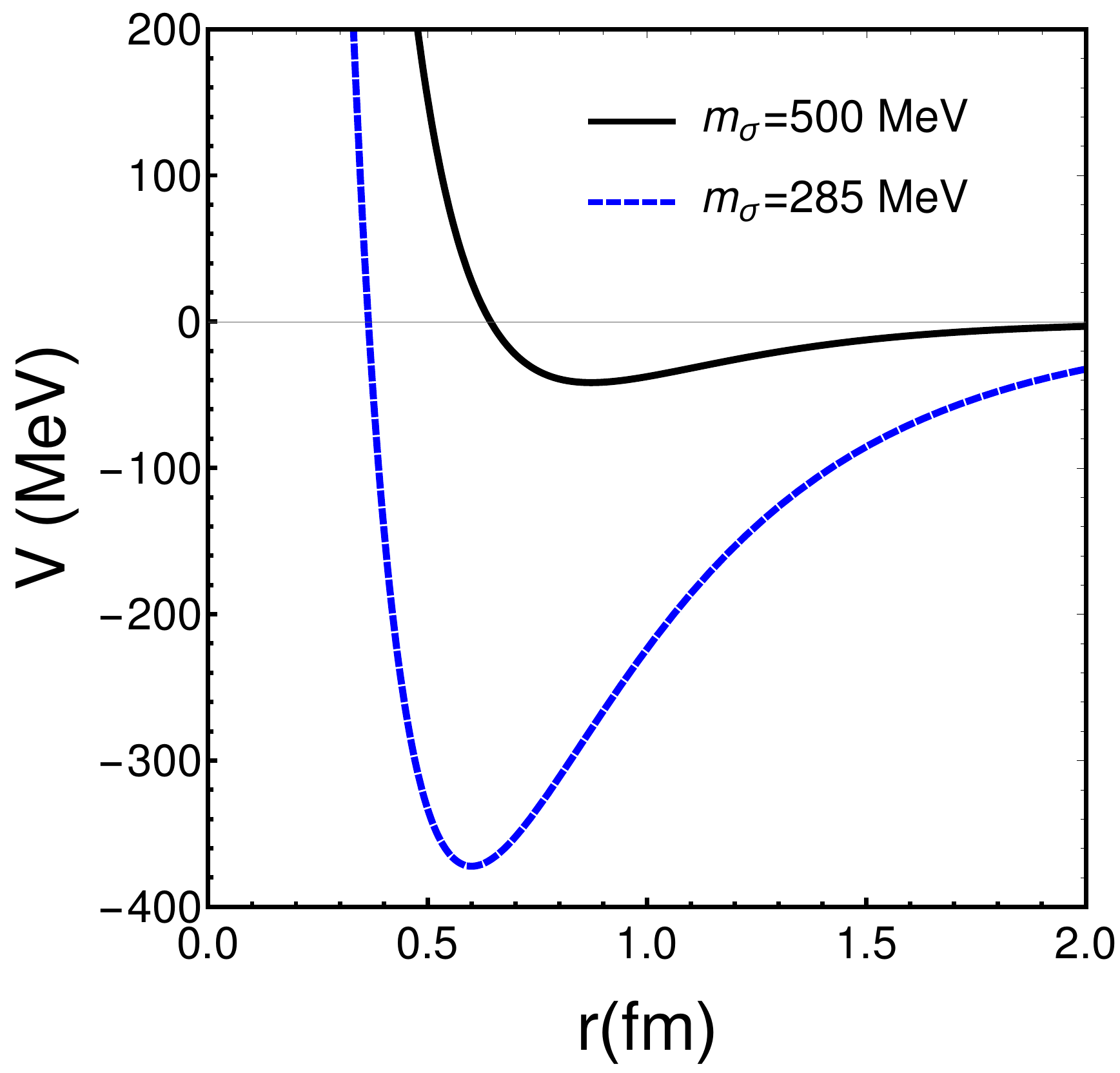}
% figure in 2N.nb
\caption{The effective nuclear potential in Eq.~(\ref{eq:NNpot}) with vacuum $\sigma$ mass (solid line) and with a reduced mass supposedly to happen close to the critical region (dashed line). Those we consider to be the minimal and maximal possible magnitude for near-critical line potential.}
\label{fig_two_potentials}
\end{center}
\end{figure}

The preclusters we discuss are statistical correlations of several nucleons at relative distances $1-2$ fm induced by the interbaryon potential $V(r)$. The phenomenological (unmodified) nuclear potential that we considered in Ref.~\cite{Shuryak:2018lgd} (called $V_{A'}$ in that paper) was the Serot-Walecka potential~\cite{Serot:1984ey},
\be \label{eq:NNpot} V(r)= - \alpha_\sigma \frac{e^{-m_\sigma r}}{r} + \alpha_\omega \frac{e^{-m_\omega r}}{r} \ , \ee 
with $\alpha_\sigma=6.04$, $m_\sigma=500$ MeV, $\alpha_\omega=15.17$, $m_\omega=782$ MeV. The potential with these parameters is shown in Fig.~\ref{fig_two_potentials} with a black solid line.

However, the relevant ratio for precluster formation is not $B/T$ but the Boltzmann factor with the \textit{maximal depth} of the potential to temperature $\exp\left(-V(r_{\textrm{min}})/T\right)$. For the potential Eq.~(\ref{eq:NNpot}) this ratio is 
\be \frac{|V(r_{ \textrm{min}})|}{T} \sim \frac{1}{3} - \frac{1}{2} \ , \nn \ee
and thus clustering is relatively scarce.

However, there are two important regimes in which this simple conclusion can be reversed, and rather large correlations can be achieved. The first regime is when the effective mass of the $\sigma$ mode is strongly reduced because of the closeness of chiral symmetry restoration at $T>T_c\approx 155$ MeV. According to studies of chiral transition at $\mu=0$~\cite{Tripolt:2013jra}, and to the discussion in our previous paper~\cite{Shuryak:2018lgd}, the initial $NN$ potential the $\sigma$ mass can be reduced from 500 MeV down to $m_\sigma \sim  285$ MeV. As shown in dashed line in Fig.~\ref{fig_two_potentials} this modification results in crucial changes of the effective potential, inverting the situation to
\be \frac{|V(r_{\textrm{min}})|}{T} \sim 2-3 \nn \ . \ee
This situation becomes more evident in the vicinity of the QCD critical point (if it exists) since the hypothetical critical mode becomes (nearly) massless, making appear long-range forces associated with its exchange.

These large Boltzmann factors $\exp( |V(r_{\textrm{min}})|/T)$ play much more important role when several nucleons are involved. For example, the system composed by $N=4$ nucleons in an approximate tetrahedral configuration has six relative potentials, so the Boltzmann factor enters with the sixth power.

In Ref.~\cite{Shuryak:2018lgd} the real-time dynamics has been quantified by means of classical Molecular Dynamics (MD) simulations. While we studied specific clusters with 4--13 nucleons, our main focus in that paper was on skewness and kurtosis of the baryon number distribution,
in connection with the results of the BES program at RHIC. We have demonstrated that even modest modifications of the nuclear potentials at the freeze-out conditions may significantly enhance the baryon correlations. 

However, the Boltzmann factor describes only classical thermodynamics, and the MD simulations account only for the classical dynamics of nucleons. In the onset of clustering, one needs to include also quantum effects, expected to reduce the formed correlations. This is the question we focus on in this paper, where we continue the study of preclusters.

More specifically, we will focus on quantum corrections of pairwise potentials between nucleons, and introduce a semiclassical method at finite temperature giving rise to the ``flucton'' configuration. This method will be also applied to states with 3 and 4 particles with a very specific symmetry. In addition, we will focus on four-nucleon preclusters of the $ppnn$ (or $\alpha$-particle) type. Only in this case one may think of all four particles as distinguishable (all in different spin-isospin states), without account for effects of Fermi-Dirac statistics. Its ground state is the only light nuclei which is relatively strongly bound. In fact, it is well known that $^{12}$C, $^{16}$O and perhaps even $^{24}$Mg have strong $\alpha$-particle correlations, and their lowest states are consistent with few $\alpha$-particle Bose-Einstein condensation~\cite{Tohsaki:2001an}. The four-nucleon preclusters are qualitatively different from two- and three-body clusters. While the latter have only one (barely) bound states, the former has one deeply-bound ground state and $\sim 50$ states incorporating next-shell excitations near zero energy. While this fact is experimentally known, it has been overlooked in any discussion (we are aware of) of the $d, t, ^3$He production. We use this novel semiclassical method, as well as the $K$-harmonics method, to correctly include quantum effects. Finally, we will comment on how experiments can access an overproduction of light nuclei, and propose new experimental measurements of light-nuclei ratios (in the same lines of the recently proposed $t p/d^2$ ratio) with an increased ability to signal the presence of the QCD critical point.

In Sec.~\ref{sec_2_nucl_Schr} we study the two- and four-nucleon systems using a genuine quantum mechanical method by solving the two-body Schr\"odinger equation and the $K$-harmonics method, respectively. We will see that important quantum corrections appear when the interaction potential dominates over thermal effects. In Sec.~\ref{sec:quantum} we introduce the ``flucton'' method at finite temperature as a semiclassical approximation to the full quantum solution. We apply it to two- and four-nucleon systems at finite temperature, and consider the effect of a modified $NN$ potential due to the critical-point dynamics. In Sec.~\ref{sec:lightnuclei} we propose several observables in the form of light-nuclei ratios in which the critical correlations could be observed in experiment. Some discussions on the connection between preclusters and light-nuclei are presented in Sec.~\ref{sec:evol}, where we also comment on the experimental situation of the $^4$He spectra and the need to account for its excited states. Finally, in Sec.~\ref{sec:summary} we present our conclusions.

%%%%%%%%%%%%%%%%%%%%%%%%%%%%%%%%%%%%%%%%%%%%%%%%%%%%%%%%%%%%%%%%%%%%%%%%%%%%%%%%%%%%%%%%%%%%%%%%%%%
%%%%%%%%%%%%%%%%%%%%%%%%%%%%%%%%%%%%%%%%%%%%%%%%%%%%%%%%%%%%%%%%%%%%%%%%%%%%%%%%%%%%%%%%%%%%%%%%%%%
%%%%%%%%%%%%%%%%%%%%%%%%%%%%%%%%%%%%%%%%%%%%%%%%%%%%%%%%%%%%%%%%%%%%%%%%%%%%%%%%%%%%%%%%%%%%%%%%%%%

\section{Theory of few-nucleon quantum systems in a thermal environment~\label{sec_2_nucl_Schr}} 

Before we begin the theory part of this paper, let us recapitulate its main goals:
\begin{enumerate}
 \item [(i)] develop the necessary tools to evaluate the density matrices for few-nucleon systems
at finite temperature;
 \item[(ii)] quantify clustering probabilities, focusing on the four-nucleon system;
 \item [(iii)] study how the clustering phenomenon depends on possible in-matter potential modifications.
\end{enumerate}

The standard textbook definition of the density matrix---the probability to find quantum/thermal system at a (multidimensional) coordinate $x_0$---is straightforward to compute as
\be P(x_0)=\sum_i |\psi_i (x_0)|^2 \ e^{-E_i/T} \label{eqn_P_standard} \ , \ee
by combining probabilities in all stationary states, bound and unbound, with their subsequent weighting with the Boltzmann factor. $\psi_i(x_0)$ and $E_i$ are the wave functions (eigenfunctions) and energies (eigenvalues) of the state $i$. We will be applying this definition in this section, first for two and then for four nucleons. An alternative semiclassical approach to this problem at finite temperature will be developed in the next section.

\subsection{The density matrix for two nucleons}

The two-nucleon problem is essentially a one-dimensional (radial) problem, so the density matrix  at finite $T$ can be calculated using a complete set of solutions of the Schr\"odinger equation. Simplifying the situation to central forces, without spin/isospin dependence and without electromagnetism, one combines the $pn,nn,pp$ pairs into one generic $NN$ case. In this case one should find with sufficient accuracy that there is one (near) bound state with essentially zero energy.

The original Serot-Walecka potential in Eq.~(\ref{eq:NNpot}), while it can lead to reasonable properties for infinite nuclear matter~\cite{Shuryak:2018lgd}, does not possess any bound state. For an illustration let us reduce the repulsion, and use $\alpha_\omega=9.42$, to increase the depth of the potential. This is the value we are using in this section only.

In addition to the $NN$ potential one needs to separately consider the centrifugal potential,
\be \Delta V^L_{\textrm{rot}}=\frac{L(L+1)}{2m_Rr^2} \ , \ee
for various nonzero values of $L=0,1,2,...$ ($m_R=m_N/2$ is the reduced mass, with $m_N$ being the nucleon mass).  In order not to deal with a continuous spectrum of scattering states we use a standard method: 
put a system  in a confining ``cup'' potential, chosen in a form 
\be V_{\textrm{cup}}=\left( \frac{r}{R_{\textrm{cup}}} \right)^8 \ , \ee
with large enough $R_{\textrm{cup}}=10$ fm.

With all these ingredients we numerically solve
\begin{align} 
  \label{eq:schrodinger} & - \frac{u_L''(r)}{2m_R} + (V_{NN}+V_{\textrm{cup}}+\Delta V^L_{\textrm{rot}}) u_L(r) =E^L u_L(r) \ , 
\end{align}
with $u_L(r)=rR_L(r)$ and the radial wave-function $R_L(r)$ has been factorized from the total one together with the spherical harmonics [$\psi({\bf r})=R(r) Y(\theta,\phi)$]. The normalization of $\psi(r)$ imposes, as usual,
\be \int dr |u_L(r)|^2= 1 \ . \ee

We find 60 energies and wave functions for each $L$. The beginning of the energy spectrum at $L=0$ is (in units of fm$^{-1}\approx 197$ MeV)
\begin{align} 
E^{L=0}_{i} = & -0.0113, 0.0749, 0.204, 0.369, 0.564, 0.786...  \nonumber
\end{align}

The only bound state is ``Walecka deuteron'' with an energy of $-2.2$ MeV and a root-mean-square (r.m.s.) radius of $\sqrt{\langle r^2 \rangle} \simeq 2.6$ fm (the physical deuteron also contains a small admixture of $L=2$ component, which we do not obtain in this simple example with a central potential).

Using this set of states one can find the quantum-thermal density matrix 
\be P(r,T)=\sum_{i,L} (2L+1) | \psi_{L,i}(r)|^2 e^{-\beta E^L_{i}} \label{P_with_l} \ , \ee
where $\beta=1/T$, and $i$ runs over all states with a given quantum number $L$. We consider $L=0,1$ and $2$ (in all our examples the angular dependence is included in the wave functions $\psi_{L,i}(r)$ and conveniently integrated over). In our approximation with an external $V_{\textrm{cup}}$ all states are bound. Otherwise, the continuum version of Eq.~(\ref{P_with_l}) should be used to account for the unbound states~\cite{Feynman_SM}.

\begin{figure}[htp]
\begin{center}
\includegraphics[width=7cm]{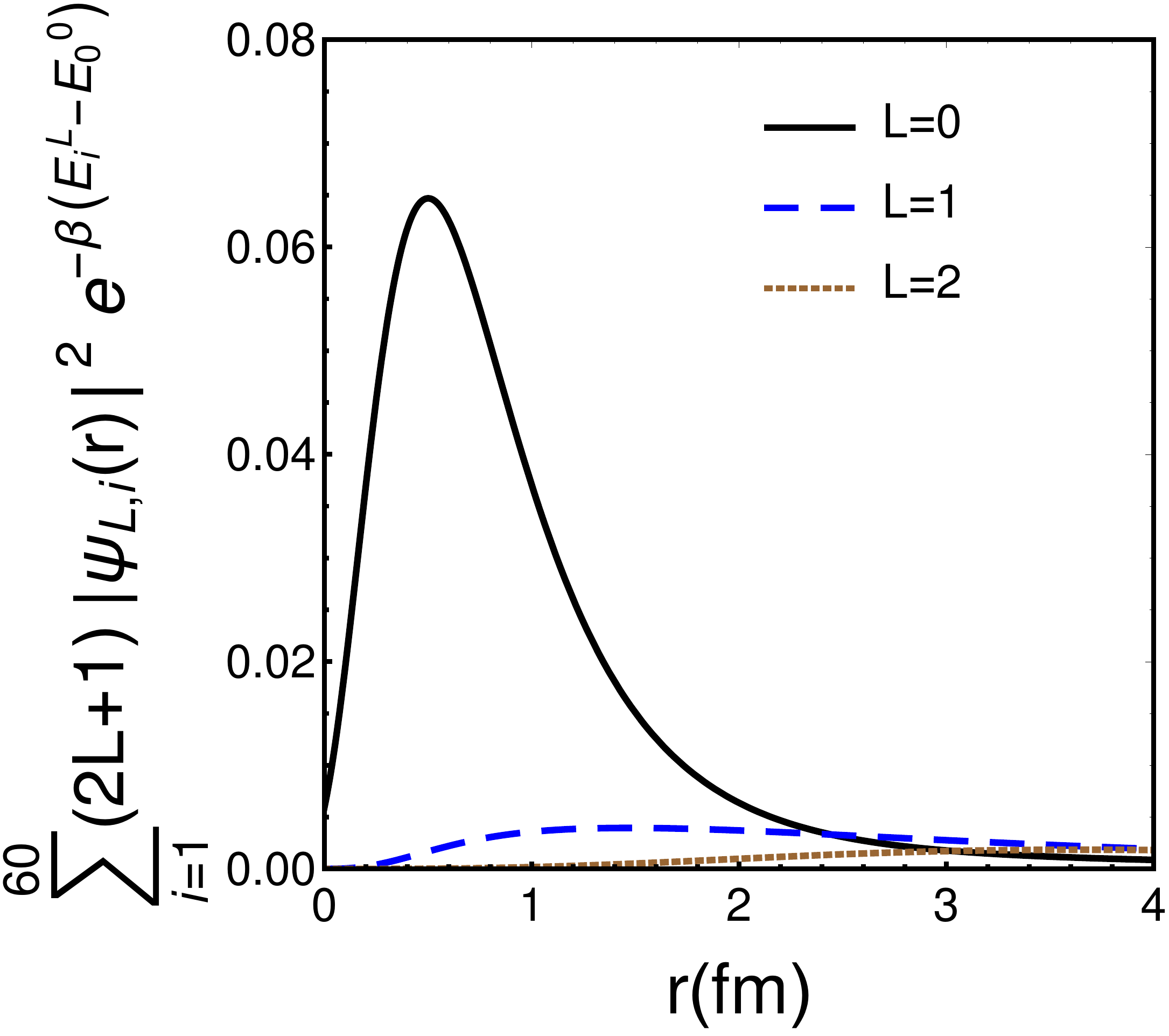}
\includegraphics[width=7cm]{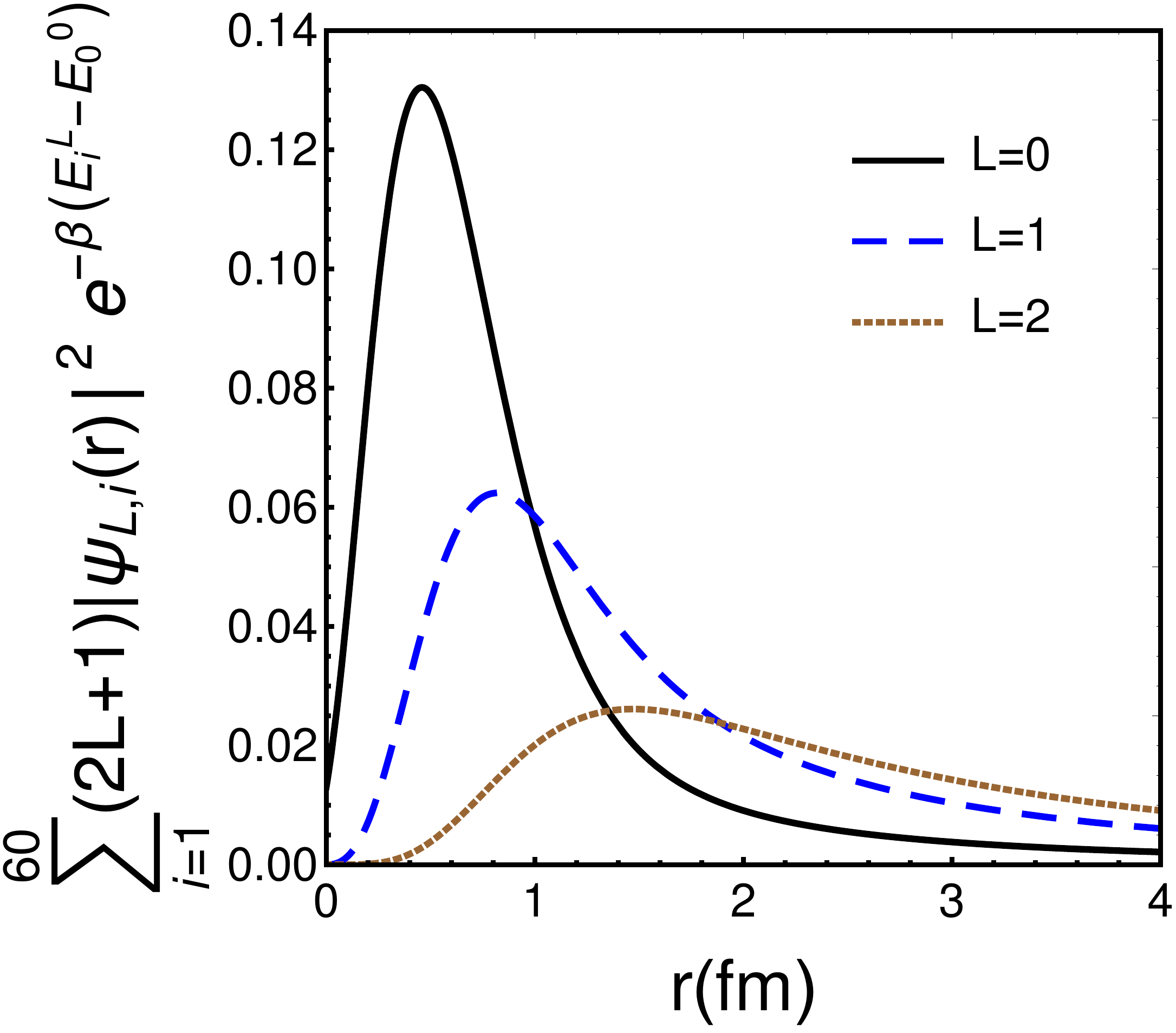}
% plots in FINALdeuteron_3.nb
\caption{The density matrix Eq.~(\ref{P_with_l}) for the Serot-Walecka potential with $m_\sigma=500$ MeV at $T = 20$ MeV (upper panel) and $T= 100$ MeV (lower panel) for different values of the quantum number $L$. The units of the $OY$ axis are fm$^{-3}$.}
\label{fig_NN_P}
\end{center}
\end{figure}

Examples at two different temperatures are shown in Fig.~\ref{fig_NN_P}, for $T \simeq 20, 100$ MeV for different angular momenta. We sum over the first 60 levels for each value of $L$. It is important to note that in the results of this figure, the $NN$ potential itself is not yet modified by the temperature. The difference between the curves is entirely given by thermal excitation of states other than the ground one.

\begin{figure}[htp]
\begin{center}
\includegraphics[width=7cm]{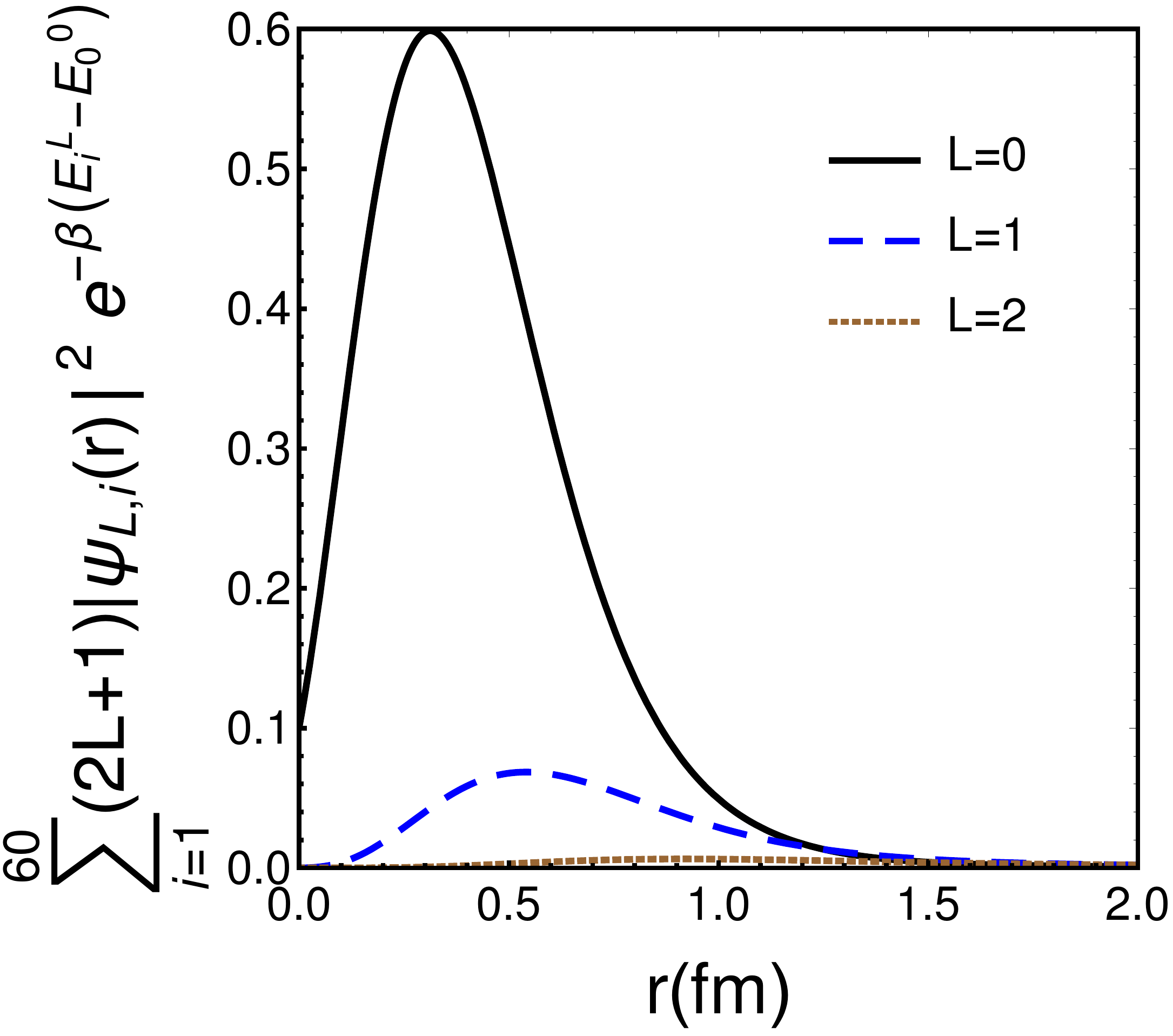}
% plots in FINALdeuteron_3modified.nb
\caption{The density matrix (\ref{P_with_l}) at $T = 100$ MeV with a medium-modified Serot-Walecka potential with $m_\sigma=285$ MeV at $T=100$ MeV for several values of $L$. The units of the $OY$ axis are fm$^{-3}$.}
\label{fig_NN_Pmod}
\end{center}
\end{figure}

From these plots one observes that states with nonzero angular momentum $L>0$ contribute only minimally at low temperatures (upper plot), even including their larger degeneracy $2L+1$. At high temperatures (lower plot) these states contribute substantially to the density matrix. However, at such temperature one also expects the in-medium modification of the $NN$ potential. Using a Serot-Walecka potential with $m_\sigma=285$ MeV we get the result at $T=100$ MeV in Fig.~\ref{fig_NN_Pmod}. Again, the higher-partial waves are subdominant with respect to $L=0$ in the density matrix. In Sec.~\ref{sec:evol} we will come back to the deuteron example and comment about the wave package interpretation of the cluster from this density matrix, and introduce the Wigner distribution of the deuteron.

\subsection{ $K$-harmonics method  and four-nucleon clusters \label{sec:4nucleon}} 

In this section we study the four-body system using a pure quantum mechanical method, the $K$-harmonics, which goes back to the 1960's~\cite{Badalian:1966wm}. Its main idea is to focus on quantum mechanics along the ``hyperdistance''  axis in the nine-dimensional space, while other coordinates can be treated via corresponding angular harmonics. We present more details in appendix~\ref{app_K_harmonics}. Historically, this method was applied only to the ground states of light nuclei, which it describes well. In particular, it correctly reproduced the binding of $^4$He~\cite{KHarmonics_he4}.

As usual, we start with the lowest, most symmetric ground states, obtained from a 1D radial Schr\"odinger equation for the hyperdistance $\rho$, defined in Eq.~(\ref{rho_definition}) as a sum over Jacobi coordinates squared. We briefly indicate in Appendix~\ref{app_K_harmonics} the derivation of the corresponding Schr\"odinger-like equation in the case of $^4$He~\cite{KHarmonics_he4} here we only note that the squared hyperdistance is related to $r$ coordinate, the distance between any two nucleons in a tetrahedral configuration, via the simple relation
\be \rho^2= \frac{6}{4} r^2 \ . \ee

Solving the eigenvalue problem in Eq.~(\ref{eqn_radial_for4}) we have obtained 40 lowest eigenstates using the simplest potential $V_1$ from Ref.~\cite{KHarmonics_he4} and the Coulomb term between the two protons. The ground state energy we find is $E_0=-27.8$ MeV,  close to the experimental value of $E_0^{\textrm{exp}}=-28.3$ MeV. 

\begin{figure}[htp]
\begin{center}
\includegraphics[width=7cm]{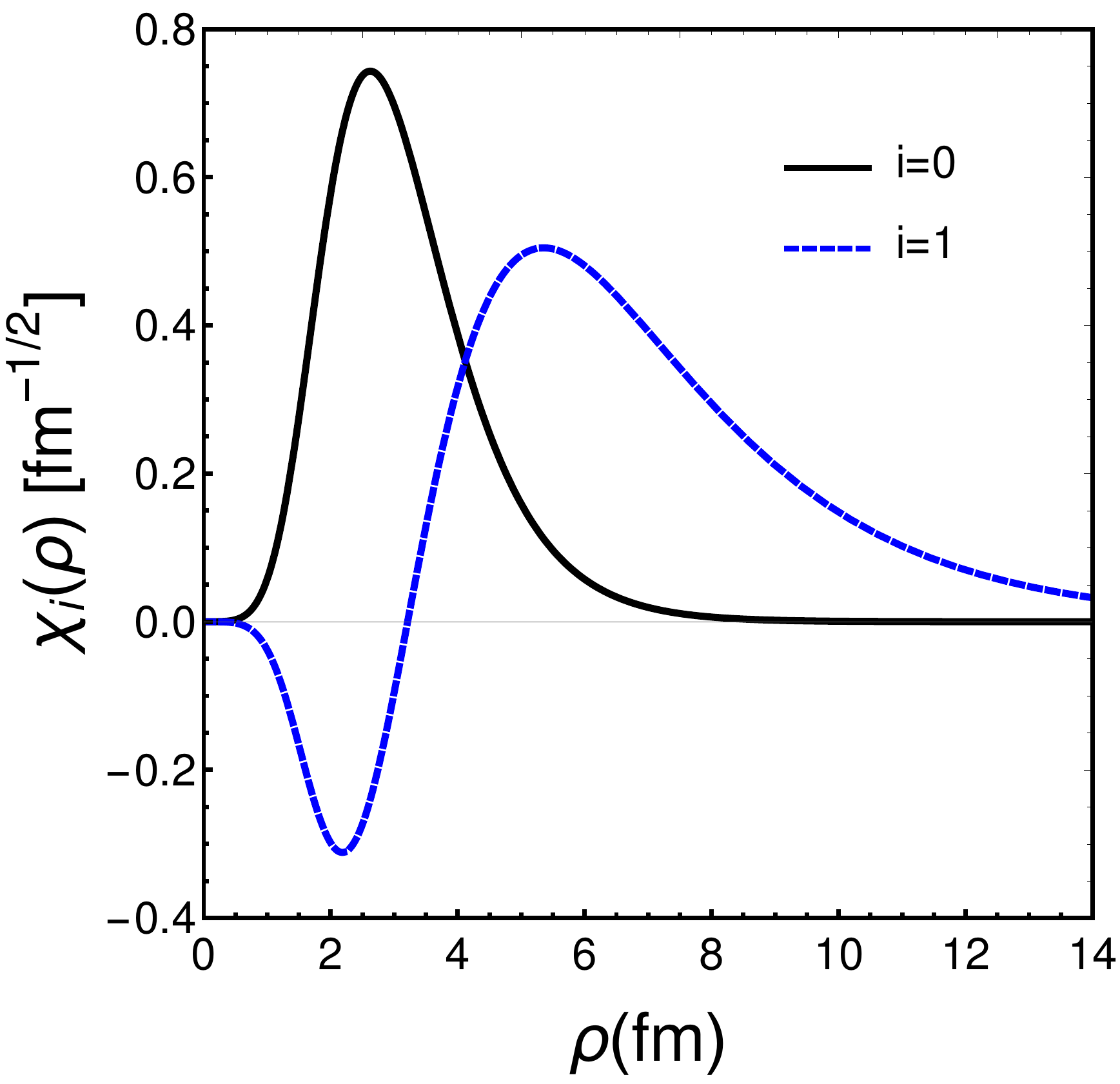}
% Plot in He4KHarmonics.nb
\caption{Two radial bound state $J^P=0^+$ wave functions for $^4$He, which are solutions of Eq.~(\ref{eqn_radial_for4}) as a function of the hyperdistance variable $\rho$. Their energies are discussed in the text.}
\label{fig_kharmonics_2wf}
\end{center}
\end{figure}

Rather unexpectedly, we also find the $second$ bound state missed by our predecessors in Ref.~\cite{KHarmonics_he4}, with $J^P=0^+$ with energy $E_1=-2.8$ MeV. To determine whether this state is physical, we show in Table~\ref{tab:he4} a compilation of the excited states of $^4$He. Among them there is just one $0^+$ state, with a binding energy of
\be B=-28.3 \textrm{ MeV}+20.2 \textrm{ MeV}=-8.1 \textrm{ MeV} \ , \ee
which is  close enough to the one we found to identify them, as the same second radial excitation. A plot with both $0^+$ wave functions $\chi_0(\rho), \chi_1(\rho)$ is shown in Fig.~\ref{fig_kharmonics_2wf}.

\begin{figure}[htp]
\begin{center}
\includegraphics[width=7cm]{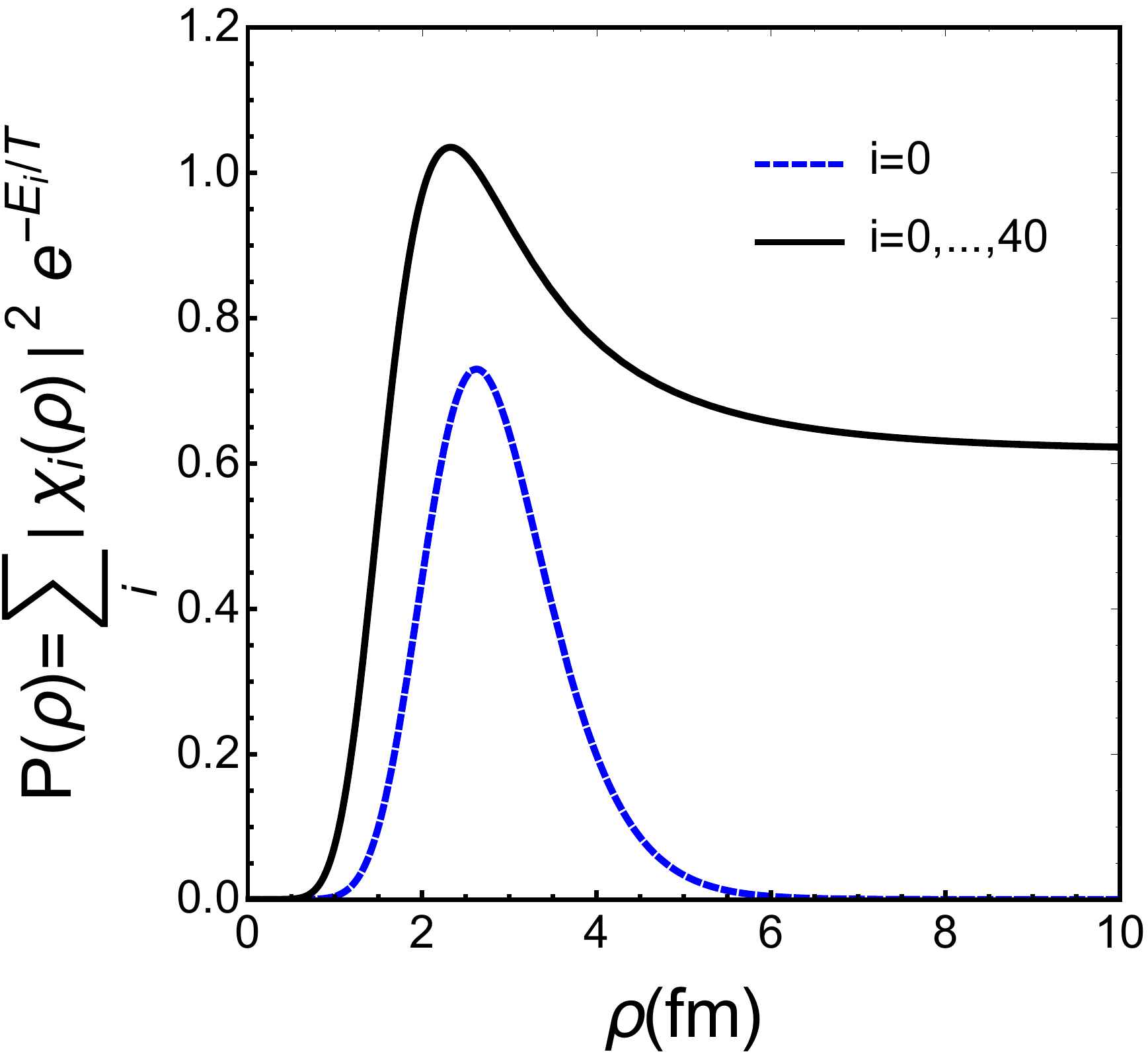}
\includegraphics[width=7cm]{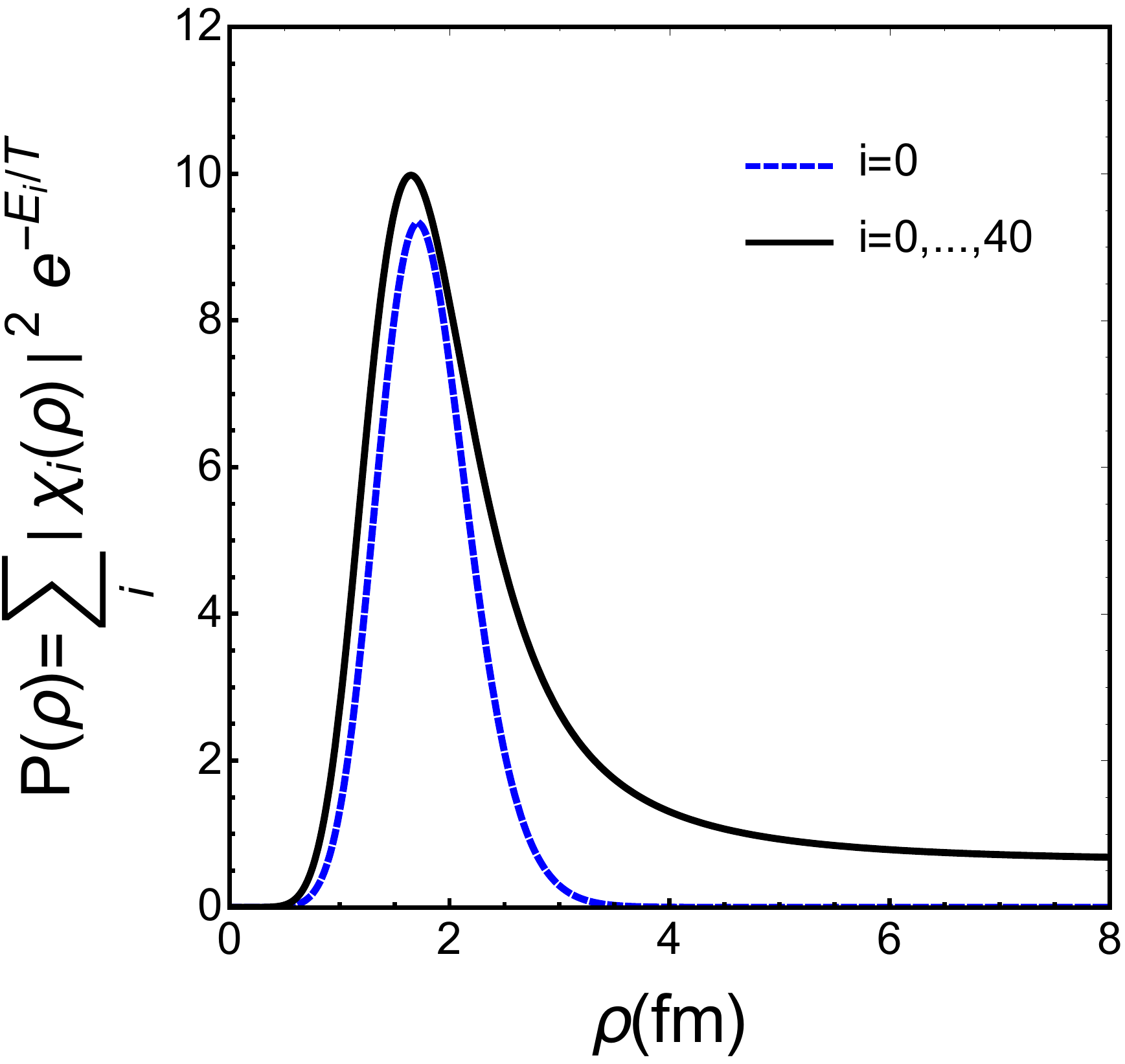}
\caption{Solid lines: Boltzmann-weighted density matrix, at $T=100$ MeV,
using 40 lowest states of the $K$-harmonics radial equation, for the unmodified nuclear potential $V_1$ used in Ref.~\cite{KHarmonics_he4} (upper plot) 
and a modified one (lower plot). In both cases the blue dashed lines show the contribution of the lowest bound state. The units of the $OY$ axis are fm$^{-1}$.}
\label{fig_kharmonics_density}
\end{center}
\end{figure}

At finite temperature, we also use the unbound states to weight them with the corresponding Boltzmann factor and calculate the thermal density matrix. The results are shown in Fig.~\ref{fig_kharmonics_density} for $T=100$ MeV. In the upper plot we present the results using the potential $V_1$ given in Ref.~\cite{KHarmonics_he4}. The solid line is the weighted density matrix at $T=100$ MeV compared to the contribution of the lowest bound state only (blue dashed line). For this (unmodified) potential the contribution of excited state to the density matrix is important as can be seen from the difference between the two curves.

\subsection{Modification of the internucleon potential}

To see what happens if the interaction potential is medium-modified, we repeat the calculation with the same form of the potential, but with the coefficient of the attractive term double. In this case the minimum of the potential reaches $\sim -400$ MeV, similarly to what happens in Fig.~\ref{fig_two_potentials}. 
 
This modified potential now has six radial bound states: Their energies in MeV are 
\be E =-226.1, -120.1,-52.6, -17.3, -3.4, -0.1 \ . \nn \ee
The corresponding density matrix  and the lowest bound state wave function squared are shown in the lower panel of Fig.~\ref{fig_kharmonics_density}. In contrast to the upper plot (for unmodified potential) the lowest state dominates the density matrix. It is not surprising (we already saw this for the $N=2$ case), since its binding is more than twice the temperature. 
In that figure we can read the magnitude of the correlation, relative to the constant asymptotic distance (the thermal contribution of propagating positive energy states) increases from $\sim 0.4$ to $\sim 12$, a huge factor.

Finally we comment about the normalization of the density matrix in the $N=4$ case. The wave function $\psi (\rho)$ in nine dimensions is normalized as
\be 1=\int |\psi_i(\rho)|^2 d^9\rho = \int |\chi_i (\rho)|^2 d\rho \ , \ee
with all the angular dependence factorized and integrated out. So the integrated density matrix has dimension of 9 or volume cube, respectively, the effect is to be multiplied by the baryon density cubed $n_B^3$. The virial expansion of statistical mechanics calls such a term the fourth virial coefficient.

%%%%%%%%%%%%%%%%%%%%%%%%%%%%%%%%%%%%%%%%%%%%%%%%%%%%%%%%%%%%%%%%%%%%%%%%%%%%%%%%%%%%%%%%%%%%%%%%%%%
%%%%%%%%%%%%%%%%%%%%%%%%%%%%%%%%%%%%%%%%%%%%%%%%%%%%%%%%%%%%%%%%%%%%%%%%%%%%%%%%%%%%%%%%%%%%%%%%%%%
%%%%%%%%%%%%%%%%%%%%%%%%%%%%%%%%%%%%%%%%%%%%%%%%%%%%%%%%%%%%%%%%%%%%%%%%%%%%%%%%%%%%%%%%%%%%%%%%%%%

\section{Semiclassical ``flucton'' method at nonzero temperatures~\label{sec:quantum}} 

In this section we introduce a novel semiclassical method to approximate the calculation of the thermal density matrix for two-, three- and four-nucleon systems. It is the generalization of the ``flucton'' path~\cite{Shuryak:1987tr} to few-body systems at finite temperature.

\subsection{Semiclassical theory at nonzero temperature}

Semiclassical approximations are well-known tools, both in quantum mechanics and quantum field theory. Standard textbooks of quantum mechanics usually start with Bohr-Sommerfeld quantization conditions, and
semiclassical Wentzel-Kramers-Brillouin (WKB) approximation for the wave function~\cite{Galindo}. Unfortunately, extending such methods beyond the one-dimensional case (or multidimensional with separable variables) proved to be difficult. Also already the first WKB correction to classical term, $1/\sqrt{p(x)}$ is not correct and contains a nonphysical singularity at the turning point.

As shown by Feynman~\cite{FH_65,Feynman_SM,kleinert2009path}, the density matrix for any quantum system can be expressed by the path integrals, over paths passing through the point $x_0$. Analytic continuation to Euclidean (Matsubara) time defined on a circle $\tau\in [0,\beta=\hbar/T]$ lead to its finite temperature generalization
\be P(x_0)=\oint \mathcal{D}x(\tau) \ e^{-S_E \left[ x(\tau) \right]/\hbar} \ , \ee
taken over the periodic paths which start and end at $x=x_0$. This expression
has led to multiple applications, perturbative (using Feynman diagrams) or numerical (e.g. lattice gauge theory).

Another interesting usage of this expression is development of a novel semiclassical theory.
Its main idea is that in certain conditions the path integral is dominated by 
 minimal action (classical)  path, called ``flucton''. The idea was introduced
in Ref.~\cite{Shuryak:1987tr} (it was also later suggested independently in Ref.~\cite{deCarvalho:1998mv}.). Unlike the WKB approximation, this approach works for multidimensional and quantum-field-theory settings. It also leads to a systematic perturbative series based on Feynman diagrams, with clear rules for each order.

Systematic application of this method at zero temperature ($\beta\rightarrow \infty$) for a number of quantum mechanical problems has been developed in Refs.~\cite{Escobar-Ruiz:2016aqv,Escobar-Ruiz:2017uhx,Shuryak:2018zji}. The quantum corrections have been calculated to three loops, and shown to be in exact agreement with asymptotic expansion of the ground-state wave functions at large distances. The reader can find all the details in these references. We present a minimal content of the ``flucton'' method in Appendix~\ref{app_fluctons}.

\begin{figure}[htbp]
\begin{center}
\includegraphics[width=7cm]{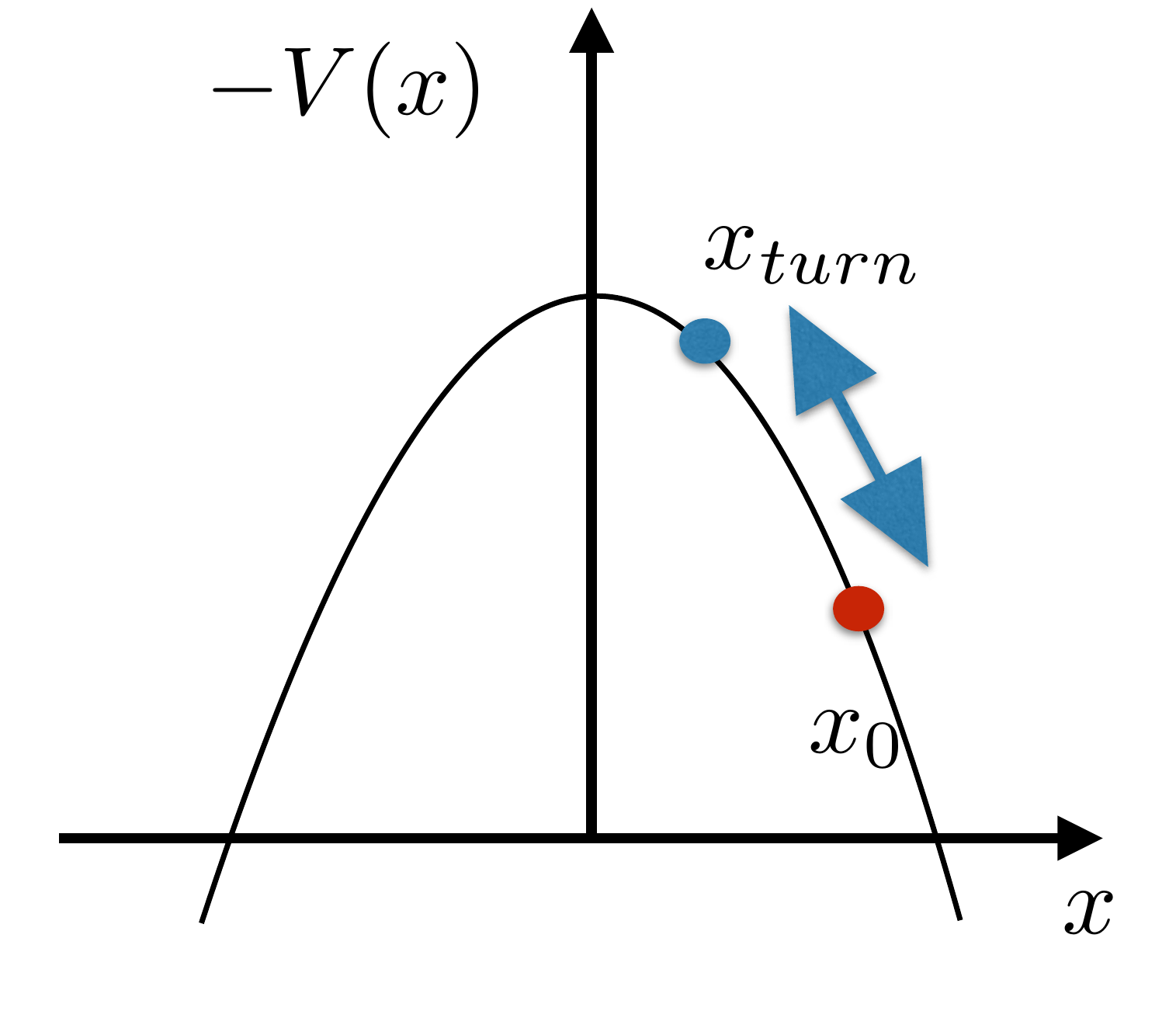}
\includegraphics[width=7cm]{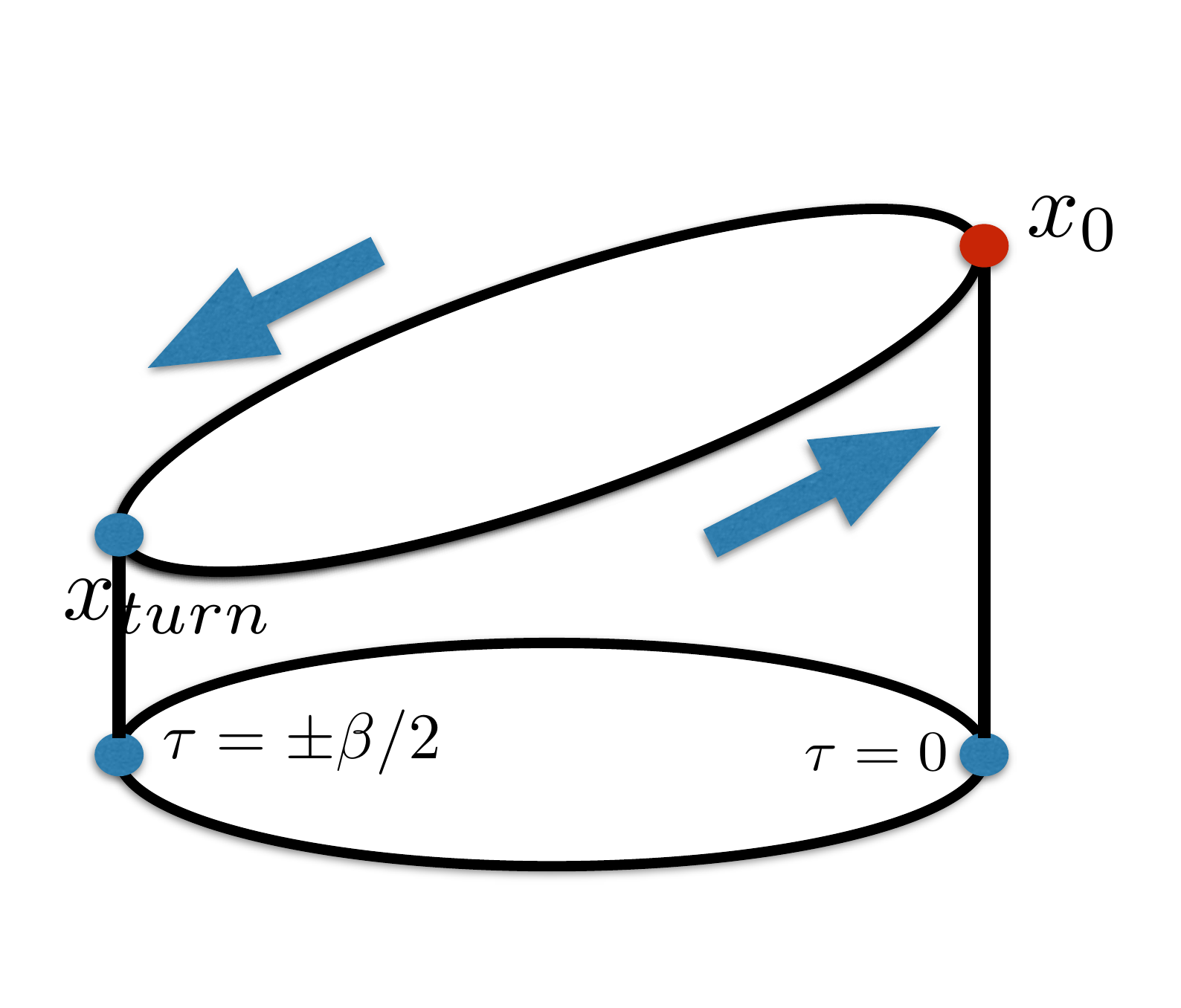}
\caption{Two sketches explaining properties of the flucton classical paths. The upper one shows the (flipped) potential $-V(x)$ versus its coordinate. The needed  path
starts from arbitrary observation point $x_0$ (red dot), goes uphill, turns back at the turning point $x_{\textrm{turn}}$ (blue dot), and returns to $x_0$  during the required period $\beta=\hbar/T$ in imaginary time.  
The lower plot illustrates the same path as a function of Euclidean time $\tau$ defined on a ``Matsubara circle'' with circumference $\beta$. }
\label{fig_fl}
\end{center}
\end{figure}

At $T=0$ quantum systems are in their ground states, and therefore studies of the density matrix are related to semiclassical description of the ground state wave functions. It has been shown in the above
mentioned papers how path integral semiclassical higher-order corrections correspond to the asymptotic expansion of solutions to Schr\"odinger equation.

At finite temperatures the path integral is modified, but it can still be dominated by certain  ``flucton'' paths, which should satisfy a number of conditions. They should
\begin{itemize} 
\item[(i)] have minimal action, thus satisfy classical equation of motion with Euclidean time $\tau=it$;
\item[(ii)] be still periodic, starting and ending at the designated observation point $x=x_0$;
\item[(iii)] have a specific time period $\beta$ in $\tau$ , the ``Matsubara  time", related to the temperature by $ \beta = \frac{\hbar}{T}$.
\end{itemize}

In Fig.\ref{fig_fl} we provide two sketches explaining how these paths look like. 

The semiclassical theory at nonzero temperature will be the subject of a separate paper~\cite{T_fluctons}. Some results, for harmonic and anharmonic oscillators, are briefly summarized in Appendix~\ref{app_fluctons}. Applications of this method to nucleon systems with $N=2,4$ are given in the following sections.

\subsection{Two-nucleon system as a thermal flucton~\label{sec_fluct_234}}

In this section we apply for the first time the flucton method described in Appendix~\ref{app_fluctons} to a two-body potential at finite temperature. Before we start let us remind two limits, in which the method leads to some obvious results:

\begin{enumerate}
\item At large $T$ ($\beta \rightarrow 0$) the periodic paths have no time to propagate, so the system stays at $x(\tau)=x_0$. The action is $S_E \rightarrow \beta V(x_0)$ which corresponds to the usual classical Boltzmann factor. 
\item At small $T$ the system is mostly in the ground state and the density matrix $P\rightarrow |\psi(x_0)|^2$. The flucton method obviously yields the semiclassical version of $\psi(x_0)$.
\end{enumerate}

For two particles the Euclidean action for their relative motion reads 
\be S_E[r(\tau)]=\int d\tau \left( \frac{m_N}{4} \dot r^2 +V(r) \right) \ , \ee
where $r$ is the internucleon distance, $\dot r=dr/d\tau$, $V(r)$ is the pairwise (inverted) potential $V_{NN}$, and the coefficient $1/4$ in the kinetic energy appears because of the use of the nucleon mass instead of the reduced mass $m_R=m_N/2$). The classical equation of motion is 
\be \ddot{r}=\frac{2}{m_N} \frac{\partial V(r)}{\partial r} \ , \ee
whose solution is the required flucton path $r(\tau)=r_{\textrm{fluc}} (\tau)$ as a function of the observation point $r_0$. The density matrix is proportional to the action of this solution,
\be P(r_0) \sim e^{-S_E[r_{\textrm{fluc}} (\tau)]} \ . \ee 
The observation point will be simply denoted as $r$ in our plots.

\begin{figure}[htp]
\begin{center}
\includegraphics[width=7cm]{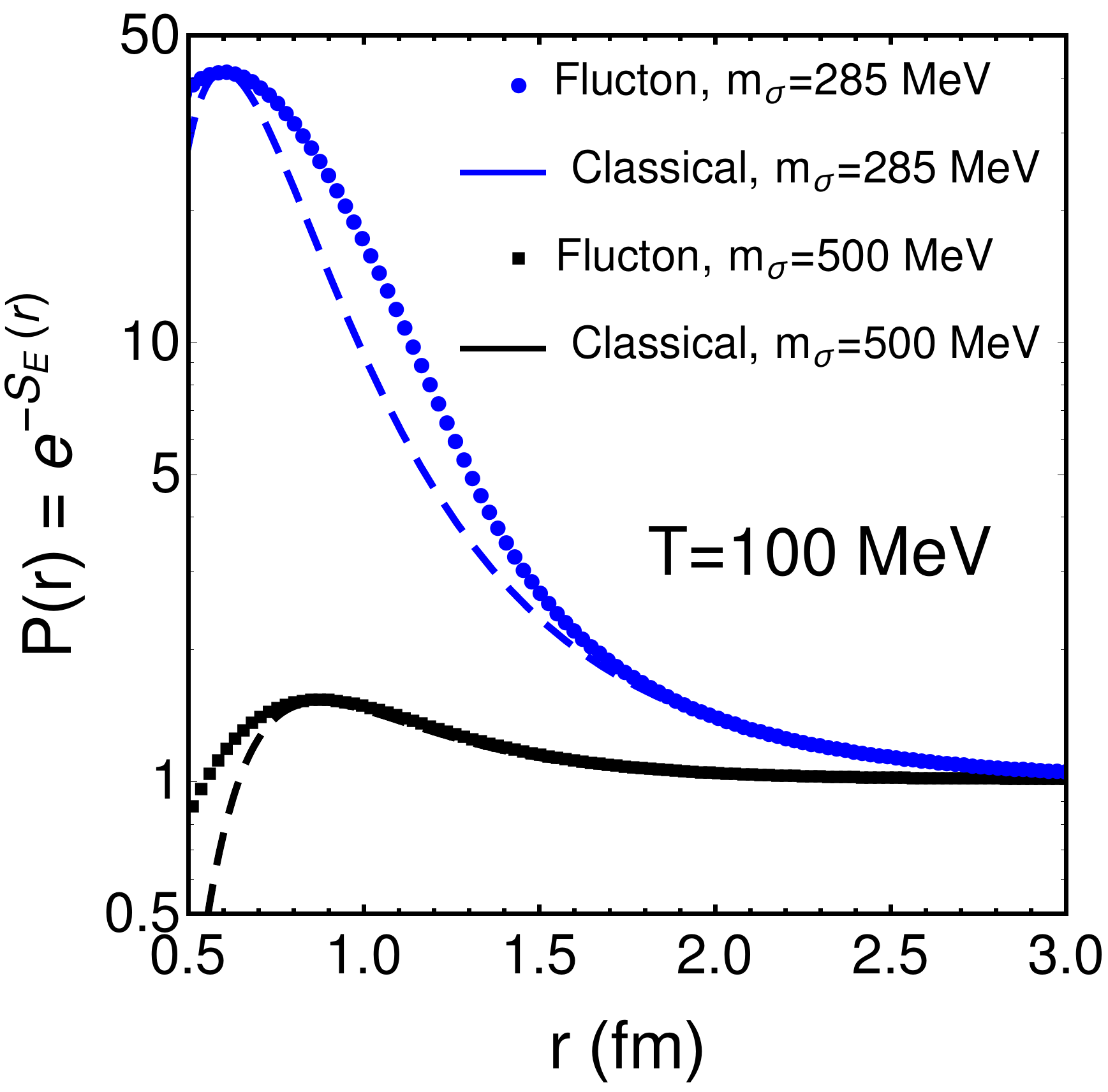}
% figure in 2N.nb
\caption{The probability for two nucleons being at distance $r$ (fm) from each other at temperature $T=100$ MeV. In symbols we plot the semiclassical probability distribution calculated via the flucton method, while lines are Boltzmann factors with the Walecka potential in Eq.~(\ref{eq:NNpot}). We used two different values of the $\sigma$ mass.}
\label{fig_2_andTwo_potentials}
\end{center}
\end{figure}

In Fig.~\ref{fig_2_andTwo_potentials} we compare the probabilities (not normalized) of two nucleons being at distance $r$ from each other at a temperature $T=100$ MeV, calculated by both methods: the flucton method (dots) and a classical Boltzmann factor (solid and dashed lines). We use two potentials, the original Walecka-like potential Eq.~(\ref{eq:NNpot}) with $m_\sigma =500$ MeV, and another with $m_\sigma=285$ MeV, with increased attraction. As one can see, for the unmodified potential the effect is rather modest, and classical thermodynamics (solid line) coincides with the semiclassical result (squares). At small values of $r$ the potential presents a steep repulsive wall, which makes the classical solution go quick to zero, whereas the flucton case presents larger probability due to the quantum barrier penetration. The comparison between methods is however very different for the modified potential, for which the correlation is significant as $V_{NN}$ it is not small compared to $T$. Again, quantum penetration into the potential to the right of the minimum makes the probability for the flucton larger than the classical expectation. This is a clear illustration of how quantum effects can be taken into account in a classical calculation.

The probability $P(r)$ is not directly normalizable. This happens because the potential has the asymptotic limit to zero when $r\rightarrow \infty$, and therefore $P(r) \rightarrow 1$.
This is similar to the pair correlation function of infinite systems (we comment on these in our Ref.~\cite{Shuryak:2018lgd}) which tends to 1 at large distances, the value of the ideal gas.
Similarly here, one should normalize $P(r)$ to the ideal gas value, e.g. to quantify the effect between potentials, we calculate the so-called correlation volume 
\be v_{\textrm{eff}}=4 \pi \int drr^2 \ [P(r)-1] \ . \ee  

For the two Boltzmann cases shown in Fig.~\ref{fig_2_andTwo_potentials}, they are $v_{\textrm{eff}}=5.3$ and $151$ fm$^3$, respectively.  The nucleon density under freeze-out conditions is a fraction of the nuclear matter density $n_0\approx 0.16$ fm$^{-3}$. Multiplying it by $v_{\textrm{eff}}$ one finds that while the original potential leads to probability of pair correlations less than one, the modified potential instead predict strong pairing of the nucleons.

\subsection{Tetrahedral thermal fluctons~\label{sec:flucN4}}

Let us now study the $N=4$ flucton case at finite temperature. To reduce the number of dimensions we will assume a particular equilibrium configuration (tetrahedron) and consider unidimensional trajectories along the mutual distance $r$. As a warm-up exercise let us work out the equations for the $N=3$ case.

For three particles we also consider a simplified configuration to reduce the difficulty of the problem. Based on symmetry, one expects that classical flucton would correspond the particles to be at the corners of a equilateral triangle, with the (time-dependent) side $r(\tau)$. Without loss of generality, this is achieved when three locations are
\begin{align} 
\left\{ \vec{x}_1,\vec{x}_2,\vec{x}_3 \right\} & =\left\{ \left( \frac{r}{\sqrt{3}},0 \right),  \left( - \frac{r}{2\sqrt{3}},\frac{r}{2} \right), \left( - \frac{r}{2\sqrt{3}},- \frac{r}{2} \right) \right\} \nn \ . \\
&
\end{align}
The length of each coordinate squared is $\vec{x}_i^2=r^2/3$ and the sum of the three adds to $r^2$, so the action is
\begin{align}
S_E & =\int d\tau \left( \sum_{i=1}^3 \frac{m_N}{2} \dot{ \vec{  x}}_i^2+ \sum_{\textrm{pairs}} V(r) \right) \nonumber \\ 
& = \int d\tau \left( \frac{m_N}{2} \dot{r}^2+3 V(r) \right) \ , \end{align}
and the classical EOM for the relative distance is
\be  \ddot{r}= \frac{3}{m_N} \frac{\partial V (r)}{\partial r} \ . \ee

In a similar manner we can directly proceed to the action and the equation of motion for four nucleons, assuming a tetrahedral shape with side $r$ (interparticle distance). 
%Starting now with  4 unit vectors, e.g.
%\be  
%  \vec e_1=(0, 0, 1), \,\,\,\,\,\, \,\,\,\,\,\,
%   \vec e_2= (2 {\sqrt{2} \over 3}, 0, -{1 \over 3}), \\
%   \vec e_3= (-{\sqrt{2}\over 3}, \sqrt{2\over   3}, -{1 \over 3}), 
%   \vec e_4= (-{\sqrt{2} \over 3}, -\sqrt{2\over 3}, -{1 \over 3}) \ee
% with $\vec e_i^2=1$ and $(\vec e_i \cdot \vec e_j)=-1/3, i\neq j$,
%  we define the (time-dependent) coordinates as $\vec x_i=\vec e_i a(t) $.
%  The kinetic energy is then 
%  \be K=\sum_i {m \over 2}\big({\partial \vec{x_i} \over \partial t}\big)^2 =4\dot{a}^2 {m \over 2}\ee
%  The potential energy in this case is $P=6 V(\sqrt{8 \over 3} a)$. Changing coordinates to
%  $R(t)=\sqrt{8 \over 3} a(t)$ one finds the corresponding action and EOM in the form
% \be S=\oint d\tau \big[{3 m \over 4} \dot{r}^2+6 V (r) \big] \ee
%  \be \ddot{R}=\big({4 \over m }\big){\partial V (R) \over \partial R} \ee
% Note that for one pair of nucleons the kinetic energy has coefficient $m/4$ because
% of reduced mass, and a single potential. So, extra factors 3 and 6 for  the quartet of nucleons 
% in the action effectively  wins only extra factor 2 in the EOM. 
In this case we have $N=4$ coordinates, which can be parametrized without loss of generality as
\begin{align}
\vec{x}_1 &=\left(0,0, \sqrt{ \frac{3}{8} } r \right), \ \vec{x}_2=\left(\frac{r}{\sqrt{3}} ,0, -\frac{r}{2\sqrt{6}} \right) , \  \nonumber \\
\vec{x}_3 &=\left(-\frac{r}{2\sqrt{3}}, \frac{r}{2}, - \frac{r}{2\sqrt{6}} \right) , \ \vec{x}_4=\left(-\frac{r}{2\sqrt{3}}, -\frac{r}{2}, - \frac{r}{2\sqrt{6}} \right) , \ 
\end{align}
with $\vec{x}_i^2=3/8r^2$.
The action and the equation of motion are,
\begin{align}
S_E & =\int d\tau \left( \sum_{i=1}^4 \frac{m_N}{2} \dot{x_i}^2+ \sum_{\textrm{pairs}} V(r) \right) \nonumber \\ 
& = \int d\tau \left( \frac{3m_N}{4} \dot{r}^2+6 V(r) \right) \ , \end{align}
and the EOM to be solved for the flucton solution is
\be  \ddot{r}= \frac{4}{m_N} \frac{\partial V (r)}{\partial r} \ . \ee

As a side remark, it is curious that the equation of motion in Euclidean time for the three cases $N=2,3,4$ follows the general expression
\be \ddot{r}= \frac{N}{m_N} \frac{\partial V (r)}{\partial r} \ . \ee
Unfortunately there are no more configurations with $N >4$ in which all particles stay at the same distance between each other so this result cannot be generalized for $N>4$ (for results with polyhedra with $N>4$ see our paper~\cite{Shuryak:2018lgd}).

\begin{figure}[htp]
\begin{center}
\includegraphics[width=7cm]{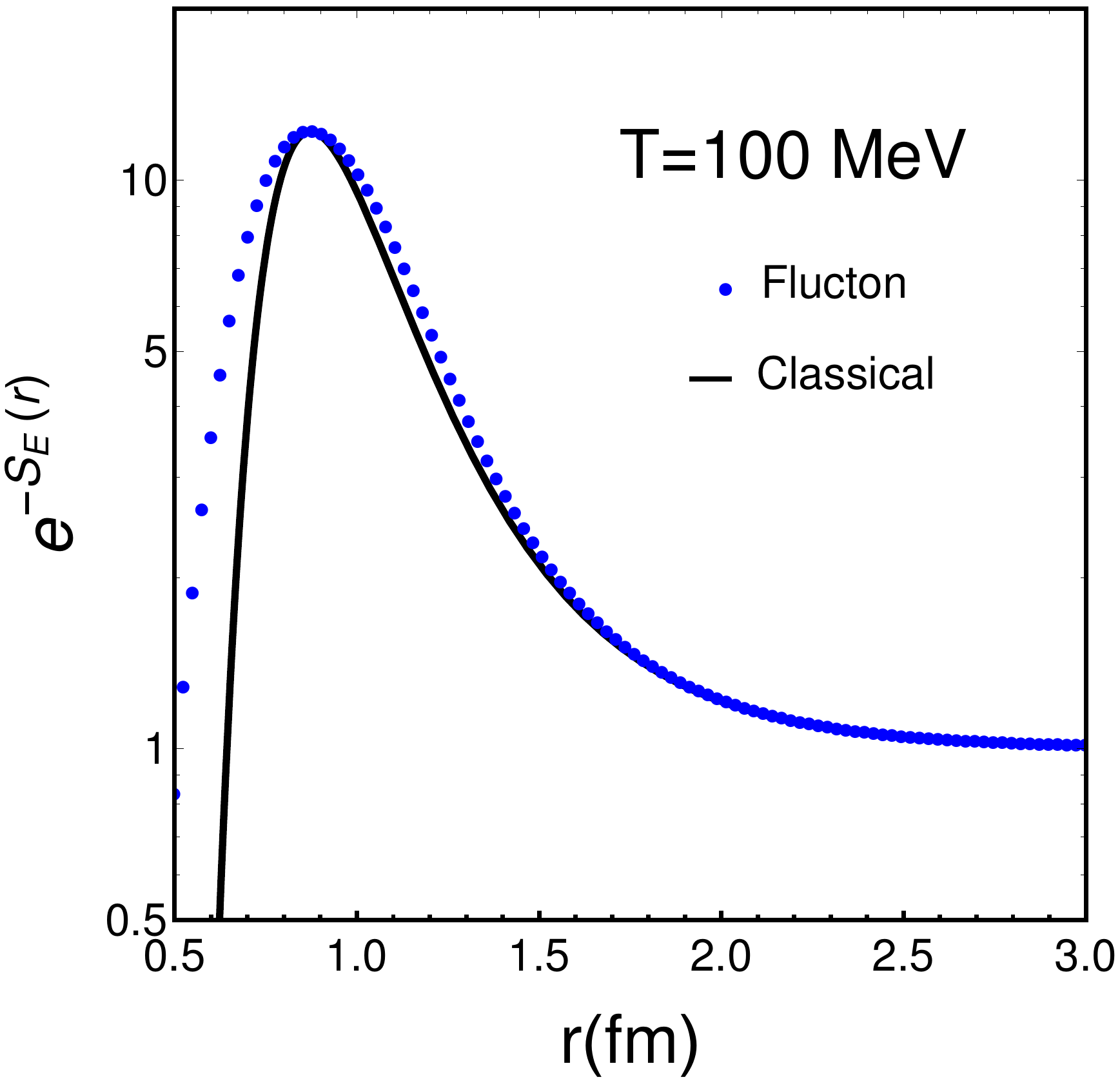}
\includegraphics[width=7cm]{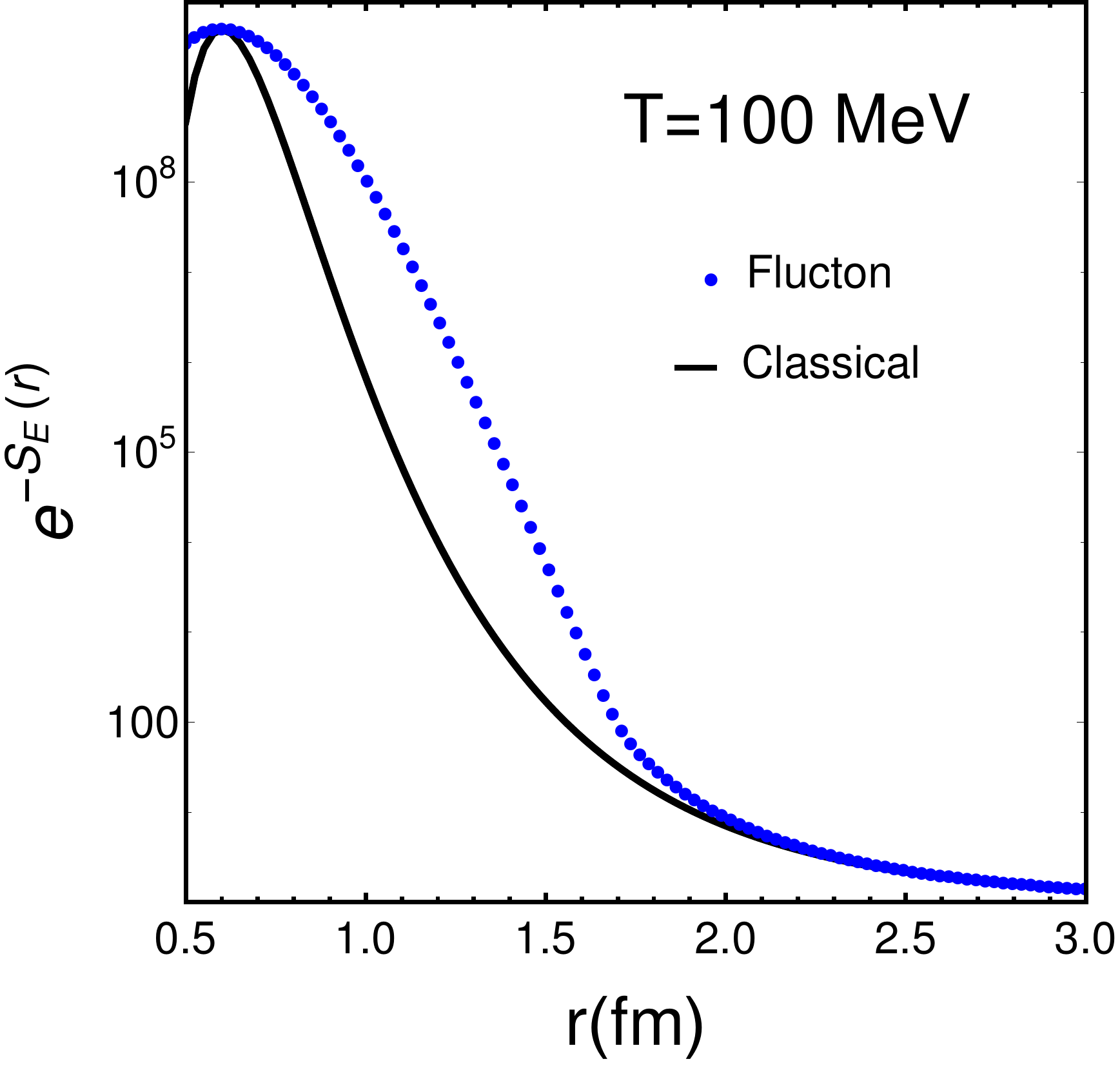}
% figure in 4N_flucJuan.nb
\caption{The exponential part of the four-nucleon density matrix $\exp (-S_E)$ calculated by the flucton method (dots) versus the classical Boltzmann exponent $\exp(-6\beta V)$ (lines) at $T=100$ MeV. The upper and lower plots are for Walecka-type potential with $\sigma$ mass $m_\sigma=500, 285$ MeV,
respectively.}
\label{fig_WalN4}
\end{center}
\end{figure}

After explaining the setup for four nucleons, we show the results of semiclassical flucton calculation, paying special attention to the sensitivity of the particular $NN$ potential used. In Fig.~\ref{fig_WalN4} we compare the semiclassical result for the density matrix and the classical Boltzmann distribution, for unmodified ($m_\sigma=500$ MeV) and strongly modified $\sigma$ meson mass ($m_\sigma=285$ MeV) in the Serot-Walecka potential.

In the former case the difference is not so large, as for the $N=2$ example, and Boltzmann expression provides a fair description. With a deeper potential the situation is quite different (see lower panel of Fig.~\ref{fig_WalN4}). Notice that the clustering is huge for the modified potential: It happens because its depth of $\sim -400$ MeV is multiplied by six pairs. Again, quantum effects (included only in the flucton solution) are important in those areas where the classical probability is suppressed.

\begin{figure}[htp]
\begin{center}
\includegraphics[width=7cm]{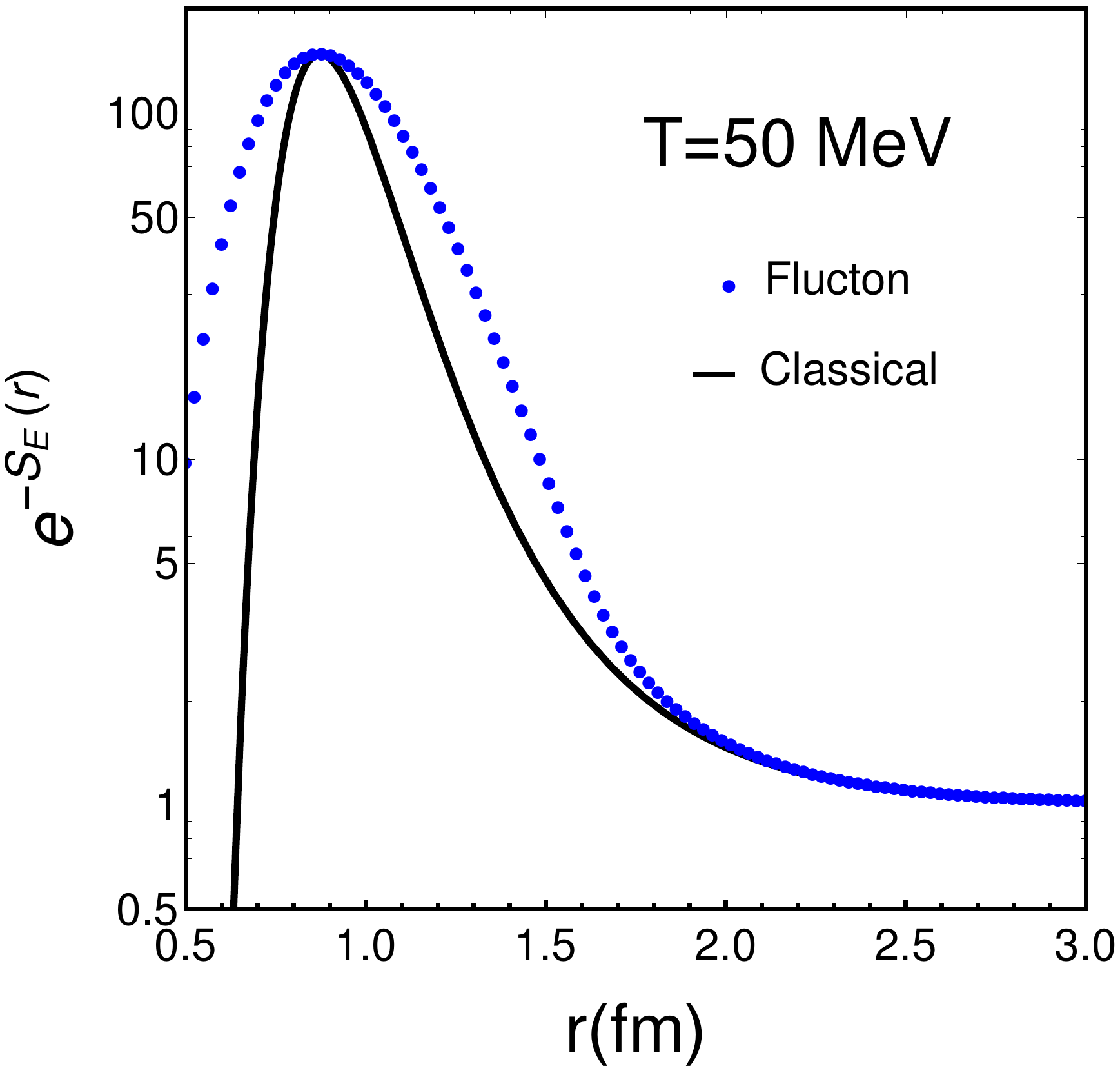}
\includegraphics[width=7cm]{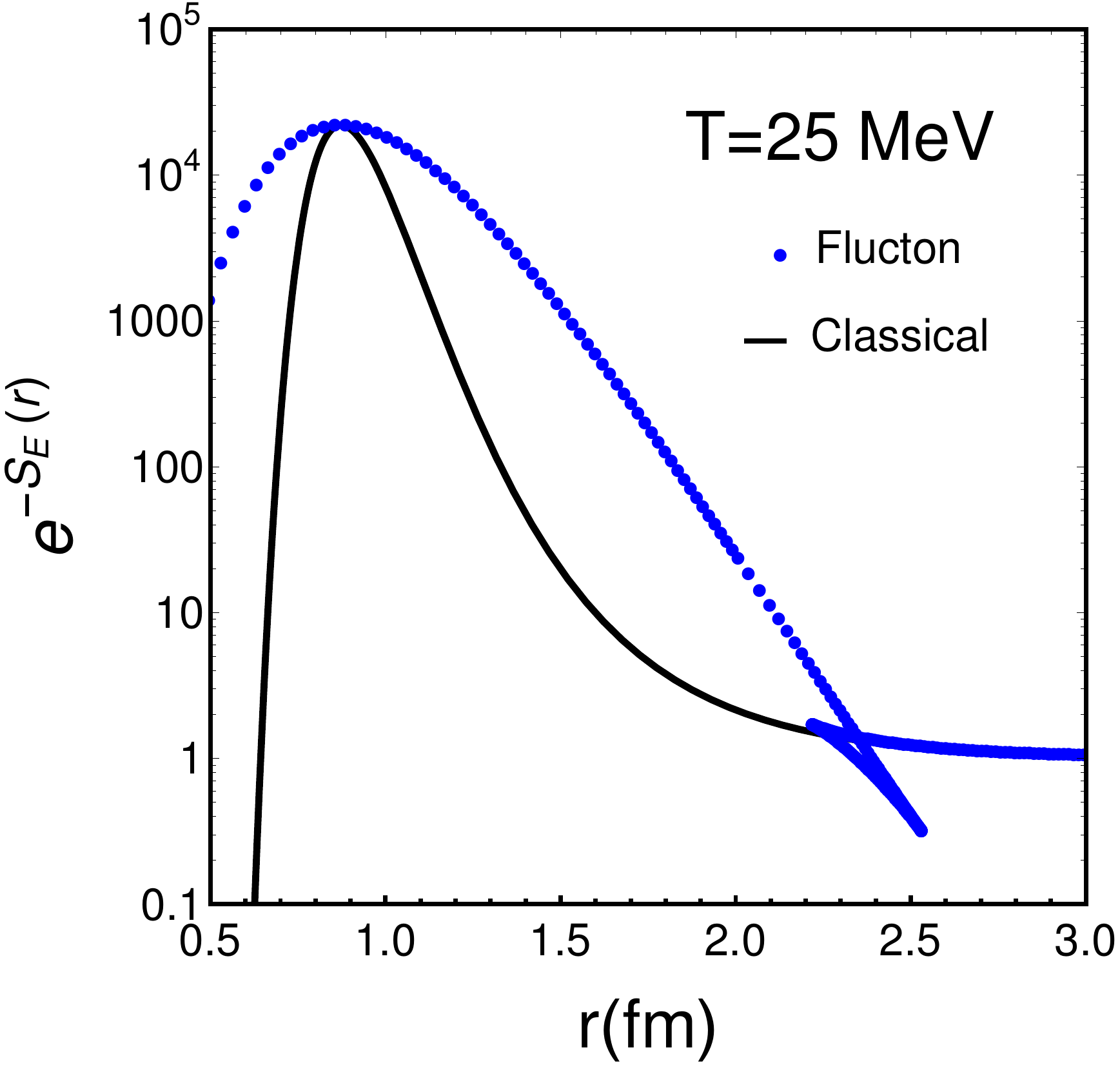}
% figure in 4N_flucJuan.nb
\caption{The density matrix  for four nucleons in the tetrahedral configuration as a function of internucleon distance $r$, for temperatures $T=50,25$ MeV, from top to bottom. The continuous lines show the classical Boltzmann factor $\exp(-6V(r)/T)$ while the dots correspond to the semiclassical flucton configuration. The $NN$ potential used is the Serot-Walecka potential Eq.~(\ref{eq:NNpot}) with $m_\sigma=500$ MeV.}
\label{fig_4N_fluct}
\end{center}
\end{figure}

To deepen a bit more the temperature dependence, in Fig.~\ref{fig_4N_fluct} we compare the exponent in the density matrix from classical statistical mechanics (solid lines) with the results of the semiclassical flucton method, for different temperatures $T=100,50,25$ MeV without modifying the potential. Note that as the temperature decreases, the quantum fluctuations make the width of the distribution significantly wider than that predicted by the Boltzmann factor.  Formally, the semiclassical approximation should be reliable when the flucton action is large, $S_E \gg1 $. In this respect the models considered in Appendix~\ref{app_fluctons}, the harmonic and anharmonic oscillators, differ from nuclear potentials. In the former cases the potential grows indefinitely away from its minimum, so the action also grows, and semiclassical approximation is improving for large distances. However nuclear potentials are short ranged, at they get small at large distances: with them $S_E$ gets small as well. As a result, semiclassical approximation is reliable only in some interval of distances.

Note the curious loop in the flucton points at $T=25$ MeV. That means that for certain values of the observation point $r_0$ (remember that in the plots the subindex 0 has been removed) the classical equations of motion provide up to three independent solutions in some region of the potential. Existence of multiple paths leading to the same final point $x_0$ and requiring the same propagation time is of course a phenomenon well known in mechanics. In fact, already in 1659 Huygens discovered the \textit{ isochrone curve},  a cycloide, sliding along which to the bottom from any initial point (at zero initial velocity) takes the same time. While several paths may satisfy the requirements needed for a ``finite temperature fluctons'' with the right period, it is not clear a priori which of these solutions should contribute  to the path integral. One could select the ones with the smallest action (the largest contribution to the path integral). In this case, all points belonging to the loop should simply be disregarded, and the semiclassical density matrix simply has a jump in the derivative, a kind of first-order transition (notice the formal similarity with the Maxwell construction for the determination of the thermodynamical potential across a first-order transition cf. Fig.~12 in Ref.~\cite{Torres-Rincon:2017zbr} by one of us). However, other solutions to the classical equation of motion might contribute as well. We plan to study this effect in detail in a future paper~\cite{T_fluctons}.

The exponent of the action, shown in Fig.~\ref{fig_4N_fluct} is huge, especially in the case of
small temperatures. The pre-exponent effects due to quantum/thermal fluctuations, not yet calculated, are expected to modify it strongly. While classical motion preserves the tetrahedral shape, quantum fluctuations do not, they happen in full $3(N-1)=9$-dimensional space and they are not scale-invariant.

\subsection{$N=4$ fluctons in the hyperdistance representation~\label{sec:fluchyper}} 

Another---and as it turns out much more realistic---approach to semiclassical theory
is to combine it with quantum mechanics along the hyperdistance $\rho$ axis.
As mentioned in the Appendix~\ref{app_K_harmonics}, it leads to appearance of effective repulsive potential $V_{\textrm{eff}} (\rho)=12/(2m_N \rho^2)$, competing with the attractive nuclear forces. 
%
%Substitution (\ref{substitution}) of the wave function leads to $\chi(\rho)$ normalized by a simple integral over $\rho$. It further results in appearance of new repulsive potential $\Delta V=12/$, which significantly suppresses the effect of attractive nuclear potential. As shown in 
%many previous works, as well as  in the previous section, 
Without it one would not be able to reproduce light nuclei binding by a simple one-dimensional equation.

It is therefore reasonable to apply the semiclassical methods, at zero or nonzero $T$, 
in the hyperdistance representation including this potential. As shown in the Appendix the effective potential for the 1D Schr\"odinger equation is (given as a function of the hyperdistance $\rho$)
\be \label{eq:poteff} V_{\textrm{eff}} (\rho)=W( \rho) +\frac{12}{2m_N\rho^2} +V_{C} (\rho)  \ , \ee
where $V_C$ is the Coulomb potential.

We solve the semiclassical equations of motion to find the flucton solution for two temperatures $T=25, 100$ MeV, using the potential $V_{\textrm{eff}} (\rho)$ and its version with a double attraction, to see the effect of the critical point on the $NN$ potential.

\begin{figure}[htp]
\begin{center}
\includegraphics[width=4.2cm]{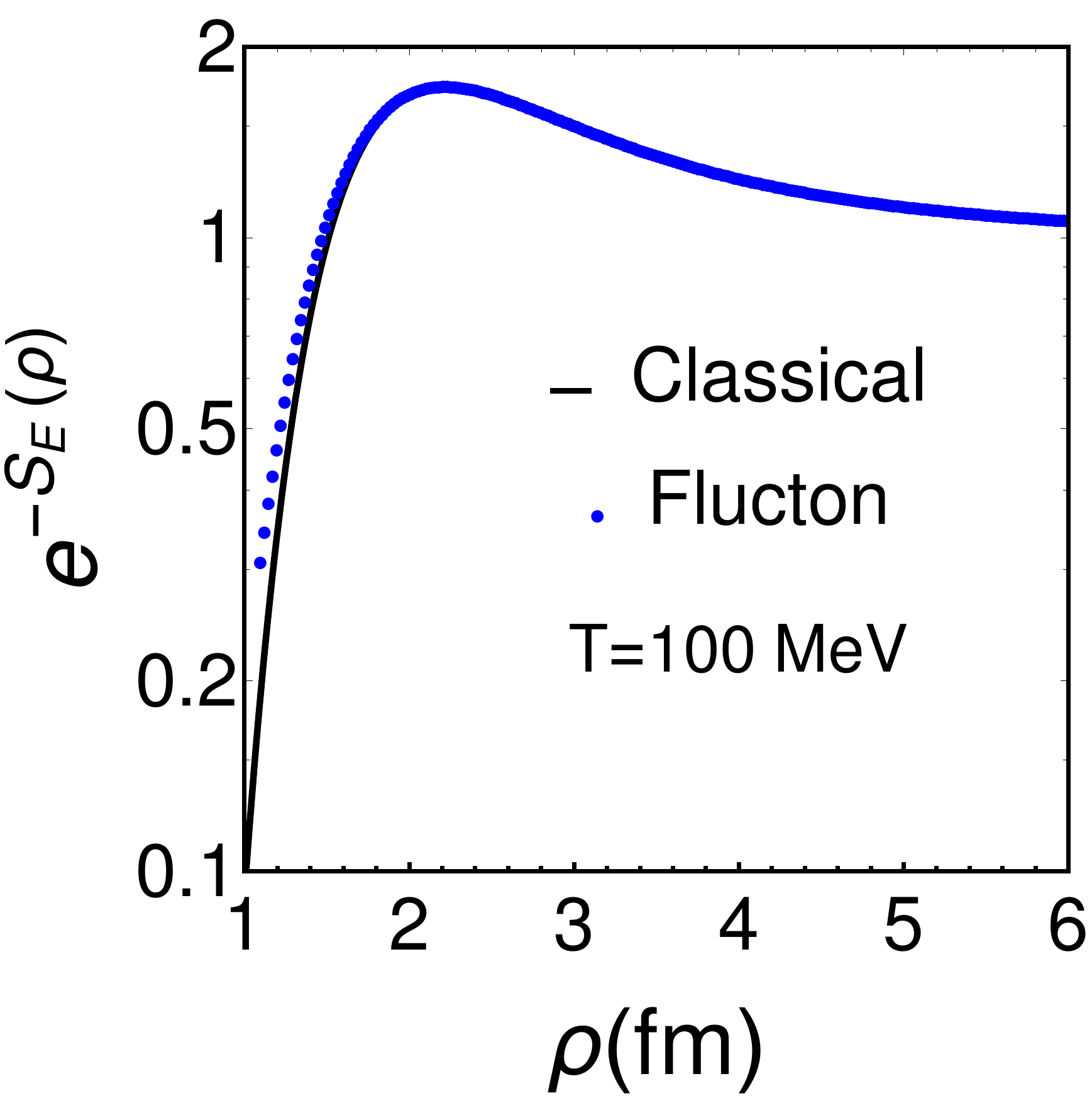}
\includegraphics[width=4.2cm]{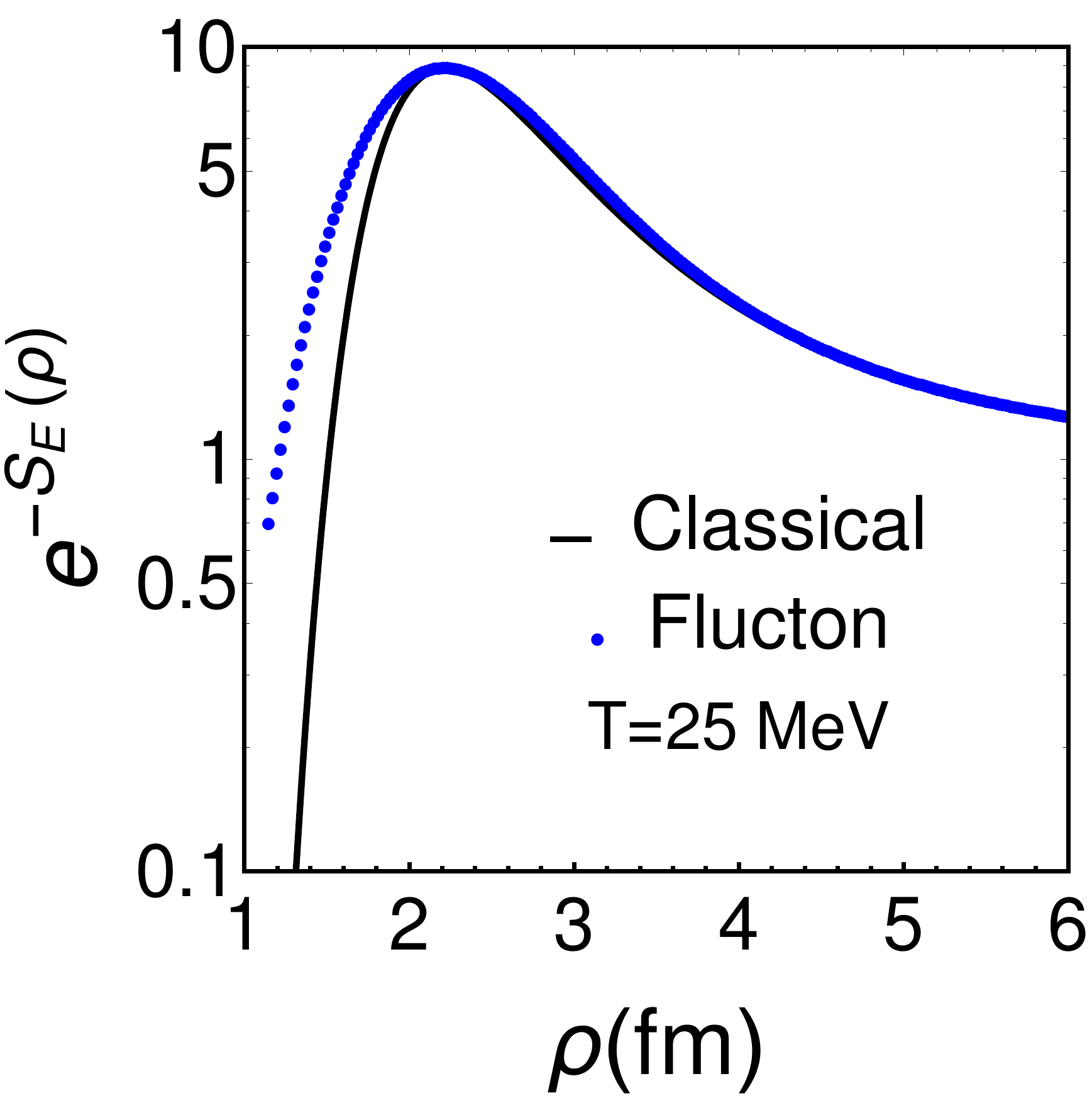}
\includegraphics[width=4.2cm]{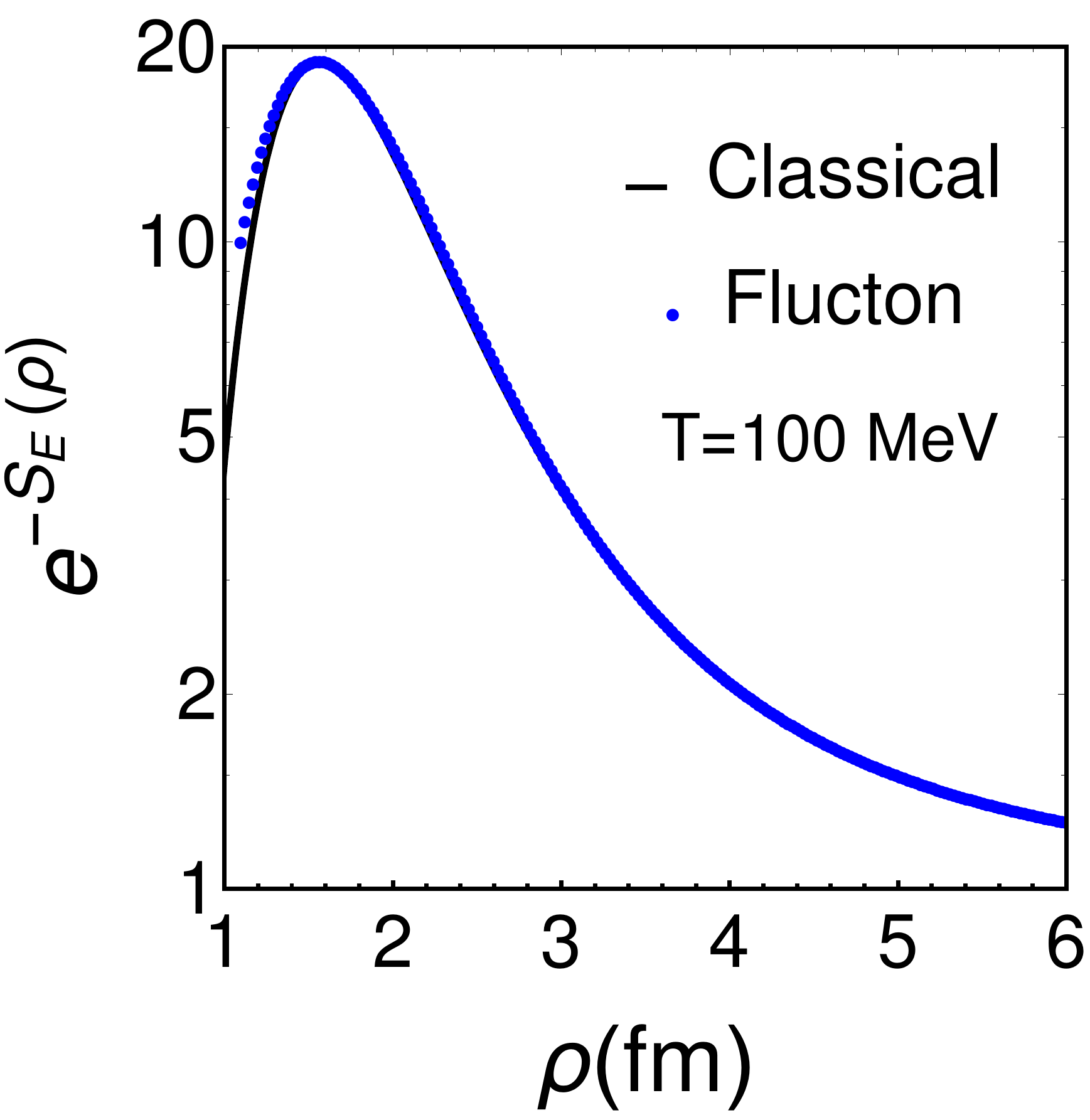}
\includegraphics[width=4.2cm]{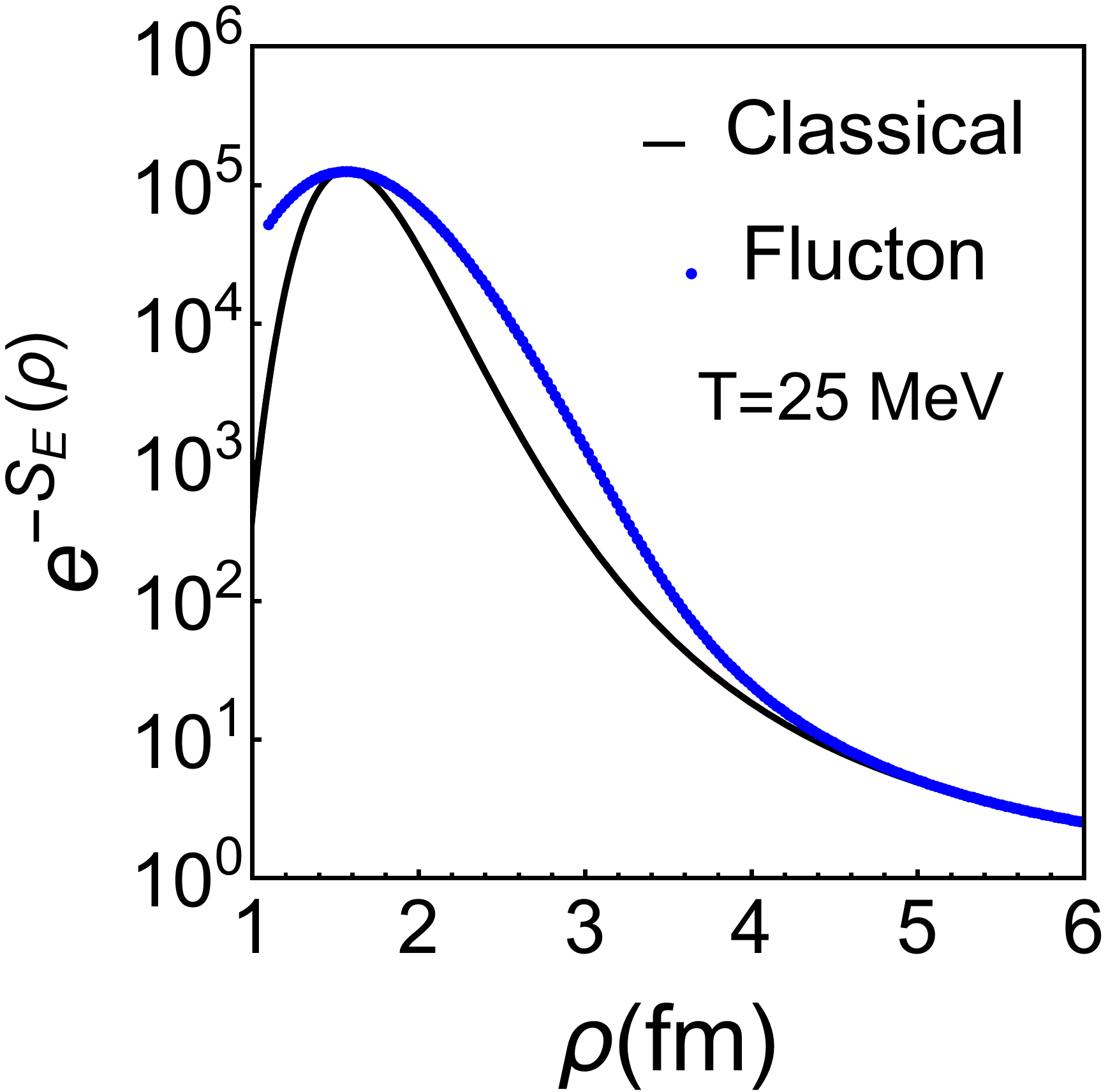}
% fig in Kharmonic.nb
\caption{$e^{-S_E}$ for the $K$-harmonics potential of $^4$He. Top panels: The $V_{\textrm{eff}} (\rho)$ used is the one shown in Eq.~(\ref{eq:poteff}) with $NN$ potential in Eq.~(\ref{eq:VNNK}). Bottom panels: same potential but with doubled attraction. Left panels: $T=100$ MeV. Right panels: $T=25$ MeV.}
\label{fig:KHarmflucton}
\end{center}
\end{figure}

In Fig.~\ref{fig:KHarmflucton} we present our results in four panels, for the four combination of temperatures and potentials.

In the top left panel we start with a temperature of $T=100$ MeV and the potential in Eq.~(\ref{eq:poteff}), where the $NN$ pairwise potential is given in Eq.~(\ref{eq:VNNK}) as used in Ref.~\cite{KHarmonics_he4}. In this case the system is classical and the Boltzmann factor account for all dynamics. For lower temperatures, where the potential is well suited, we can see the result in the top right panel. One can already see some quantum deviations from Boltzmann expectations. In the bottom left panel we present the $NN$ potential Eq.~~(\ref{eq:VNNK}) with increased attraction by a factor of 2 at $T=100$ MeV. In this example the temperature is still dominating over the potential, and a sizable deviation from Boltzmann is only obtained for $T=25$ MeV.

%%%%%%%%%%%%%%%%%%%%%%%%%%%%%%%%%%%%%%%%%%%%%%%%%%%%%%%%%%%%%%%%%%%%%%%%%%%%%%%%%%%%%%%%%%%%%%%%%%%
%%%%%%%%%%%%%%%%%%%%%%%%%%%%%%%%%%%%%%%%%%%%%%%%%%%%%%%%%%%%%%%%%%%%%%%%%%%%%%%%%%%%%%%%%%%%%%%%%%%
%%%%%%%%%%%%%%%%%%%%%%%%%%%%%%%%%%%%%%%%%%%%%%%%%%%%%%%%%%%%%%%%%%%%%%%%%%%%%%%%%%%%%%%%%%%%%%%%%%%

\section{Preclusters and production of light nuclei~\label{sec:lightnuclei}}

The main motivation for this paper is that the possible modification of the $NN$ potential close to the freeze-out time, will lead to a preclustering effect of nucleons in heavy-ion collisions. Furthermore, this effect is significantly enhanced if the internucleon potential is modified due to the $\sigma$ mass modification near the QCD critical point~\cite{Shuryak:2018lgd}.
 
Before proceeding to discuss potential observables, let us start by reminding once more what we call the preclustering phenomenon. It is very important to keep in mind that the preclusters we study are 
very different from ``nuclear fragments'', and also light nuclei (cf. Table I in Ref.~\cite{Shuryak:2018lgd}).
 
The light nuclei we will be discussing, with $N=2,3$ nucleons, $d,t, _\Lambda^3$H,$^3$He, typically have only one bound state. Furthermore, they all have very small binding energies, even in nuclear standards. The deuteron binding is only $B_d=2.2$ MeV. An extreme case is the hypertriton $_\Lambda^3$H$=pn\Lambda$~\cite{Adam:2015yta}: its binding energy is only~\cite{Juric:1973zq}\footnote{A recent measurement by STAR collaboration gives a value three times larger~\cite{Adam:2019phl}.}
\be B_\Lambda(^3_\Lambda \textrm{H})=0.13 \pm 0.05 \textrm{ MeV} \ . \ee
Clearly the physical sizes of these states are very large, comparable or larger than fireballs they come from.  

These objects are therefore very fragile, easily destroyed in any collision due to large cross section, and the cascade codes typically predicted strong suppression of their production. And yet, as shown in Ref.~\cite{Andronic:2017pug}, their production rate is in good agreement with the prediction of statistical thermal model based on ``resonance gas'' thermodynamics.  This model knows only vacuum masses of these particles, entirely ignoring their small binding. To reconcile the data with codes, in literature~\cite{Oliinychenko:2018ugs} some so-far unobserved ``resonances'' were introduced, which have small sizes and ``reasonable'' destruction cross section, decaying into light nuclei after freezeout. The explanation we suggest is that one does not need such hypothetical resonances: their role is played by \textit{ preclusters} we study. They are not bound states or resonances, just statistical correlations, with an energy uncertainty $\Delta E\sim T$ and relatively compact in coordinate space. 

Basically, there are two experimental signatures of preclusters. One, discussed in detail in Ref.~\cite{Shuryak:2018lgd} is a modified proton multiplicity distribution. Another one, which
we will address below, is a certain modification of light-nuclei production.

As we already mentioned, overall production of light nuclei (and antinuclei) is well reproduced
by the statistical thermal model; see, e.g., Ref.~\cite{Andronic:2017pug}. By ``overall'' we mean that each extra nucleon (antinucleon, upper sign) is suppressed by the same factor $\exp[-(m_N \pm \mu_B)/T_{ch}]$. The fitted values of chemical freezeout temperature and baryon chemical potential are key parameters, which give us ideas about matter as enters the hadronic world, and their dependence on the collision energy is well documented in~\cite{Andronic:2017pug}. From the experimental results of the NA49 collaboration~\cite{Anticic:2016ckv} one can also see the good agreement between the $^3$He and $t$ multiplicity and the thermal model at different collision energies. However, recent preliminary results of STAR collaboration~\cite{Zhang:2019wun} do not show a similar agreement. It will be important to study in the future the origin of this discrepancy with the thermal model.

However, behind this (overall successful) description one can observe some ``finer structure''.
It becomes visible in ratios, where the mentioned suppression factors cancel out.
One observable ratio is the {\em tritium-proton-deuterium} combination defined as
\be \label{eq:triton} {\cal O}_{tpd} = \frac{N_t N_p}{N_d^2} \ , \ee
has been previously discussed in Ref.~\cite{Sun:2017xrx}. 

In this work we also propose the following ratios involving $^4$He $(=\alpha)$
\be   {\cal O}_{\alpha p ^3\textrm{He} d} = \frac{N_{\alpha} N_p}{N_{^3\textrm{He}} N_d} \ , \quad  {\cal O}_{\alpha tp^3\textrm{He}d} = \frac{N_{\alpha} N_t N_p^2}{N_{^3\textrm{He}} N_d^3} \ . \ee
All these ratios have the same powers of fugacity in denominators and numerators, which thus 
cancel, eliminating the trivial dependence on baryonic chemical potential.
Furthermore, in classical statistical mechanics the momentum and coordinate partition functions factorize, simplifying the discussion. Mean kinetic energy per nucleon, either
a single one or inside any precluster, is the same, $\langle K\rangle =3T/2$. So, in all ratios
the kinetic parts of the Boltzmann factor, $\exp(-K/T)$ for each nucleon, do cancel as well. Volume factors also cancel. What is left are factors from statistical weights,  powers of masses in the preexponent, and potential energies:
\be {\cal O}_{tpd} =  \frac{4}{9} \left( \frac{3}{4} \right)^{3/2} \frac{\langle e^{-3V/T}\rangle_t}{\langle e^{-V/T}\rangle_d^2 }\approx 0.29  \langle e^{-V/T}\rangle \ , \label{O_tpd}
\ee
where the factor 3 in the exponential reminds that in tritium there are three nucleon pairs, and the
right-hand side is simplified under approximation that the \textit{ averaged} relative potential is the same.
Analogously,
\begin{align} 
 {\cal O}_{\alpha p ^3\textrm{He} d} & = \frac{1}{3} \left( \frac{2}{3} \right)^{3/2} \frac{ \langle e^{-6V/T}\rangle_{\alpha}}{\langle e^{-3V/T}\rangle_{^3\textrm{He}}  \langle e^{-V/T}\rangle_d } \nonumber \\
& \approx 0.18  \langle e^{-2V/T}\rangle \ , \label{eq:ratio2}
\end{align}
where $6$ is the number of nucleon pairs in $^4$He, and
\begin{align}
 {\cal O}_{\alpha t p ^3\textrm{He} d} & = \frac{8}{54} 2^{3/2} \frac{ \langle e^{-6V/T}\rangle_{\alpha}  \langle e^{-3V/T}\rangle_{t}}{\langle e^{-3V/T} \rangle_{^3\textrm{He}} \langle e^{-V/T}\rangle_d } \nonumber \\
& \approx 0.42  \langle e^{-3V/T}\rangle  \label{eq:ratio3} \ .
\end{align}

Related to this last example, if one has an approximate isospin symmetry, then one can also consider the simpler ratio
\begin{align}
 {\cal O}_{\alpha p d} & = \frac{N_\alpha N_p^2}{N_d^3} = \frac{4}{27} 2^{-3/2} e^{\mu_Q/T} \frac{ \langle e^{-6V/T}\rangle_\alpha }{\langle e^{-V/T} \rangle_{d}^3}  \nonumber \\
& \approx 0.05 e^{\mu_Q/T}  \langle e^{-3V/T}\rangle \ ,  \label{eq:ratio4}
\end{align}
where $\mu_Q$ is the charge chemical potential signaling a possible breaking of the isospin symmetry. Notice that STAR collaboration has performed statistical thermal fits in the BES completely neglecting this chemical potential~\cite{Adamczyk:2017iwn}, whereas NA49 collaboration has extracted this parameter in their fits getting values $\mu_Q/T \simeq -0.05$ MeV~\cite{Anticic:2016ckv}, so one can safely neglect it in what follows.

After introducing all these ratios let us look at experimental results of the first of them, ${\cal O}_{tpd}$.

\begin{figure}[ht]
\begin{center}
\includegraphics[width=8.6cm]{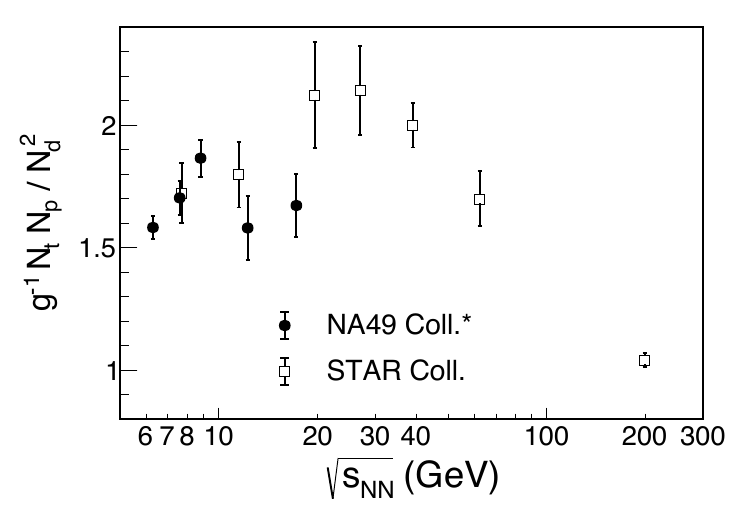}
\caption{The ratio Eq.~(\ref{eq:triton}) as a function of collision energy. The ratio is normalized by the corresponding statistical weight $g=0.29$. Note that the high-energy RHIC point at the right side of the plot gives the ratio value consistent with 1. Deviation from 1 is related to nonzero interaction potential as shown in Eq.~(\ref{O_tpd}).}
\label{fig_ratio_tpdd}
\end{center}
\end{figure}

In Fig.~\ref{fig_ratio_tpdd} we show available experimental data on the energy dependence of the combination Eq.~(\ref{O_tpd}), normalized by relevant statistical weights in $g=0.29$. Ignoring the $t$ and $d$ bindings in a statistical model, one would expect this combination to be equal to unit value. It is indeed the case at $\sqrt{s_{NN}}=200$ GeV (the most-right point), with good accuracy. 

Focusing on the specific ratios of $\alpha,t,p,d$ production, in which many kinematical factors drop out, one should expect their non monotonous energy dependence. The status of experimental measurements of these ratios is as follows. A maximum in $t-p-d$ combination was originally reported, by NA49~\cite{Anticic:2016ckv}, to be around $\sqrt{s_{NN}}\approx 9$ GeV. Very recent preliminary data~\cite{Liu:2019ppd} from STAR BES also see a maximum, although at $\sqrt{s_{NN}} =20-30$ GeV.

However, at collision energies $\sqrt{s_{NN}}\sim 10-40$ GeV the  value of the ratio observed is larger than one, roughly by the  factor 2. If correct, then this enhancement implies that under such conditions the potential and the temperature are comparable $V/T\sim \mathcal{O}(1)$ as indicated in the relation Eq.~(\ref{O_tpd}). We suggest that this extra tritium production comes from the preclustering phenomenon we discuss.

With the current data accuracy it is not possible to tell whether Fig.~\ref{fig_ratio_tpdd} show a one-maximum or a double-hump distribution. Let us note, that apart from the hypothetical
QCD critical point, the non monotonous behavior can be caused by the onset of other (perhaps less exciting but still very important) phenomena that are also expected in the same energy range. 

One of them is the \textit{ maximum fireball lifetime} as a function of $\sqrt{s_{NN}} $, well documented by recent femtoscopy data 
\cite{Lacey:2014wqa}, located  at $\sqrt{s_{NN}} \approx 47$ GeV. As indicated already on the early study~\cite{Hung:1994eq}  of  hydrodynamical expansion,
there are  two reasons for its existence, playing together in this energy range. Those are: {\bf (i)} the ``softest point'' in the equation of state, a minimum in the speed of sound $c_s^2=(dP/d\epsilon)_s$ or maximal compressibility of matter. {\bf (ii)} the maximal re-scattering rate at the freeze-out. When the densities of pions and nucleons are comparable $N_\pi\sim N_N$, the largest relevant cross section (reaching $\sigma_{\pi N} \sim 200$ mb at the $\Delta$ peak) is most effective. \\
%(iii) the last by not least, a hypothetical critical point, if it occurs there,  
% should modify the potentials and enhance what we call preclustering.

Focusing only on STAR data, and assuming that the deviation from 1 and the corresponding peak of the ${\cal O}_{tpd}$ ratio is due to the modification of the $NN$ potential, it is very tantalizing to consider the ratios Eqs.~(\ref{eq:ratio2}), (\ref{eq:ratio3}), and (\ref{eq:ratio4}), as heavier nuclei (with a larger number of nucleon pairings) would produce an enhanced effect. These ratios involving $^4$He would increase the power in the exponential by a factor 2 or 3.

\begin{figure}[ht]
\begin{center}
\includegraphics[width=7.5cm]{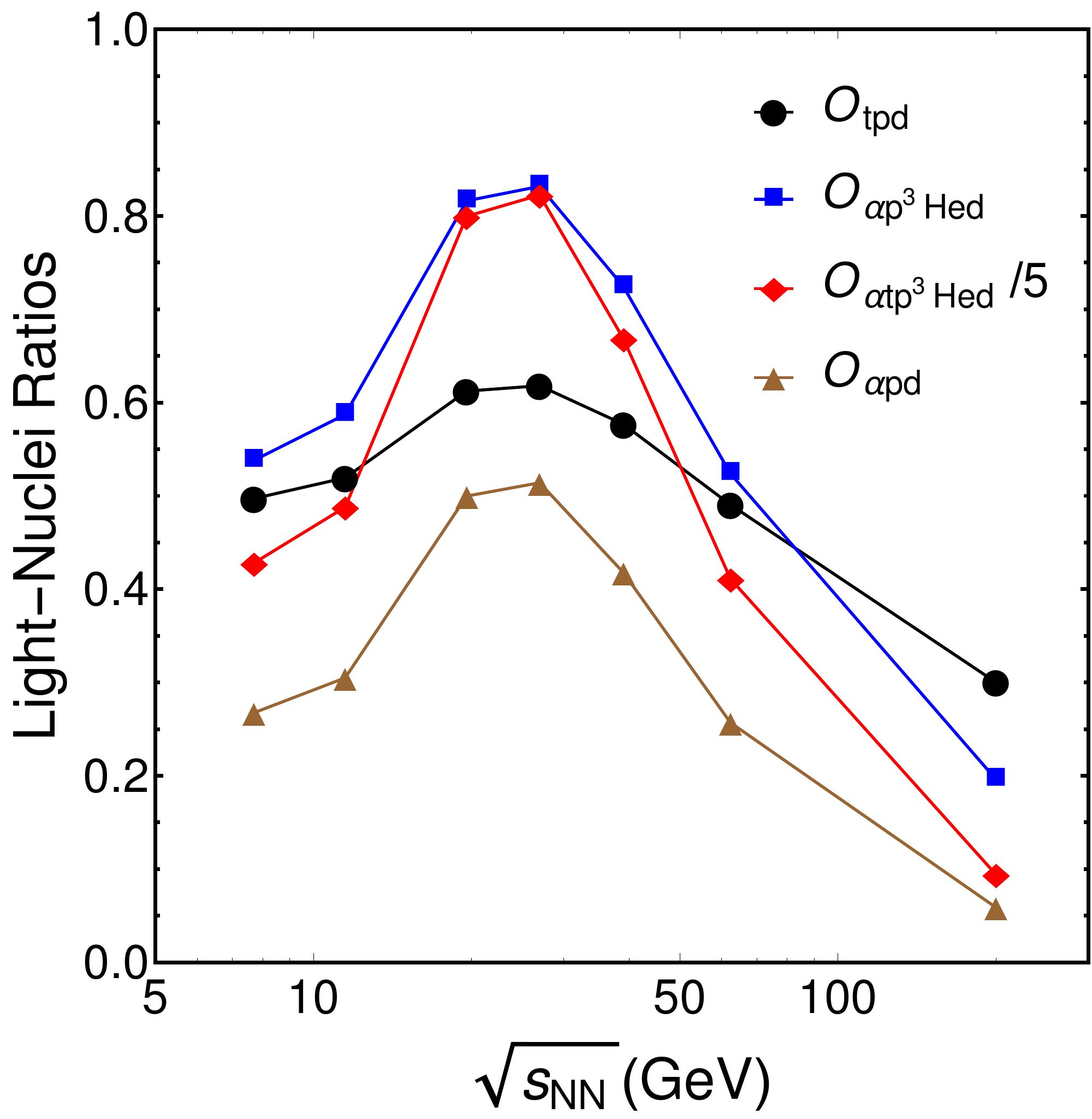}
\caption{The ratios (\ref{eq:ratio2},\ref{eq:ratio3},\ref{eq:ratio4}) as a function of the collision energy computed from the ratio (\ref{O_tpd}) in~\cite{Liu:2019ppd} assuming the only effect of the $NN$ potential modification.}
\label{fig_otherratios}
\end{center}
\end{figure}

Assuming the effect is entirely ascribed to the modification of $V_{NN}$, it is very easy to generate an approximate prediction for each of these ratios, using experimental ratio ${\cal O}_{tpd}$~\cite{Liu:2019ppd} as input. We plot the results in Fig.~\ref{fig_otherratios} for each of the ratios (notice that the result for $ {\cal O}_{\alpha t p ^3\textrm{He} d} $ has been divided a factor of 5). 
While the absolute value of these ratios depend on spin degeneracies and other factor, the important feature is the relative difference between the peak and the values at low and high energies.

If the experimental reconstruction of $\alpha$ particles~\cite{Acharya:2017bso} can be performed in these low-energy collisions, and their multiplicity measured with certain level of confidence, then these ratios would prove the sensitivity of the $NN$ potential to the presence of a near-by critical point.

It is important to mention that on top of the purely thermal production, the particle yields suffer from feed-down of hadron resonances which must be taken into account, being the most relevant to this study those decaying into protons. While we have not considered this effect, we assume that this proton feed-down is constant enough around the critical region, so that a maximum of these ratios can still be sensible indicators of the critical behavior.

%%%%%%%%%%%%%%%%%%%%%%%%%%%%%%%%%%%%%%%%%%%%%%%%%%%%%%%%%%%%%%%%%%%%%%%%%%%%%%%%%%%%%%%%%%%%%%%%%%%
%%%%%%%%%%%%%%%%%%%%%%%%%%%%%%%%%%%%%%%%%%%%%%%%%%%%%%%%%%%%%%%%%%%%%%%%%%%%%%%%%%%%%%%%%%%%%%%%%%%
%%%%%%%%%%%%%%%%%%%%%%%%%%%%%%%%%%%%%%%%%%%%%%%%%%%%%%%%%%%%%%%%%%%%%%%%%%%%%%%%%%%%%%%%%%%%%%%%%%%

\section{From preclusters to light nuclei~\label{sec:evol}}

The understanding of the formation of various nuclear species is among the central topics of nuclear physics, extensively studied in cosmological and astrophysical settings. As commented in the introduction, what is common to the regimes in which nuclei are produced in cosmos is that the available temperatures are much \textit{ lower} than the binding energies, $T \ll |B|$. The nuclear binding therefore dominates the respective Boltzmann factors  $\exp(-B/T)$. 

The setting we discuss here---the freeze-out of high-energy heavy-ion collisions---is in the opposite regime, in which light-nuclei bindings are few MeV and negligible, $B \ll T$. One might therefore think that such fragile objects cannot be produced. In other words, ``Snow flakes do not jump out of a hot oven.'' We already mentioned that experiments show this conclusion to be wrong, and we now propose and explanation.

There are basically two ideas which we try to develop in this work. One is that this pessimistic conclusion does not hold for systems of four and more nuclei. First of all, the ground state binding of $^4$He is no longer small, even for unmodified potential. Second, starting with four-nucleon clusters, multiple ($\sim 50$) states exist near zero binding, with interesting decay modes. Furthermore, let us emphasize, once again, that one should not be looking at the stationary states or their binding, but for preclustering phenomenon.  Therefore, one has to compare $3T/2$ (the average thermal energy per nucleon) to the value of the \textit{total potential energy} per particle,
\be V_i = \frac{1}{2} \sum_{j\neq i} V_{ij} \ \nn \ee 
produced at the location of particle $i$ by all other particles. For four particles there are
three terms in the sum, and even for an unmodified potential at 1 fm distance $3 |V(1 \textrm{ fm})|\sim 100$ MeV, comparable to $T$. For a modified potential like the one shown by blue line in Fig.~\ref{fig_two_potentials}, the value is one order of magnitude larger than $T$. For an increasing number $N$ of particles $V_i \sim (N-1)/2 \langle V(r) \rangle \gg 3T/2$, and the corresponding Boltzmann factor would lead to very a strong clustering. Of course, this argument does not hold for very large $N$ with the standard nuclear potential, because due to its short-range nature, nucleons start to become blind to those far away from them. Nevertheless the values of $N$ when this happen increase with the criticality of the potential, as it becomes more long-ranged.

Potential deviations of the nuclear ratios from  the statistical predictions imply that interaction strength $V$ and $T$ are comparable, in the specific collision energy range. This is only possible when the distances between nucleons are 1--2 fm$/c$, which we called ``preclusters''~\cite{Shuryak:2018lgd}. Our dynamical studies in Ref.~\cite{Shuryak:2018lgd} have shown that the corresponding correlations can be large, especially if the nuclear forces are modified as expected.

In this section we comment on the differences between the precluster formation and the final, observable, light-nuclei production. While the former at produced in the hot regime $B \ll T$, where a potential modification of the nuclear interactions are expected, the later are only observed in a situation with vanishing temperature where the standard $NN$ potential dominates the nuclear dynamics.

\subsection{Precluster decay into  stationary states}

When discussing preclusters we have so far calculated the thermal density matrix in coordinate space $P(x;T)$; see, e.g., Figs.~\ref{fig_kharmonics_density}, \ref{fig_2_andTwo_potentials}, and \ref{fig_WalN4}. This function typically has the form of a peak, centered at distances $\sim 1$ fm between particles (or hyperdistance $\rho\sim 2$ fm) tending to a constant at large distances. Let us introduce the notion of \textit{ precluster wave package}, which by definition is proportional to the square root of the density peak in coordinate density matrix
\be |\psi_{\textrm{cluster}}(x)|^2 \sim \left( P(x) - P(x=\infty) \right) \ . \ee

Because the asymptotic value at large distances is subtracted, this wave package is by definition well localized. For the $^4$He case, this would be the wave package in which four nucleons are at freeze-out. When the thermal medium rapidly disappears after that, this precluster wave package evolves further. Its decomposition into stationary states $|\Psi_n\rangle$, with the appropriate phases, 
\begin{align} 
\psi_{\textrm{cluster}}(t,x) & = \sum_n \langle \Psi_n(x) | \psi_{\textrm{cluster}}(0,x) \rangle e^{-i E_n t} \Psi_n(x) 
\end{align}
takes a time $\Delta t \sim \hbar/\Delta E$, where $\Delta E\sim E_{n+1}-E_n$ is the level spacing.
As we will see, for the excited states of $^4$He this $\Delta E$ is of the order of few MeV, so this decomposition takes a long time, much longer than explosion itself. Therefore, there is no paradox of ``fireball creating objects larger than itself'': the stationary states (with large sizes) do appear much later in time, basically at zero density! Furthermore, these states are also unstable and decay into smaller systems: This also takes a similarly long time  $\Delta t \sim 1/\Gamma\sim 50$ fm (see further discussion in Sec.~\ref{sec:exictedhe4}).

\subsection{On Wigner function projection}

If one would like to refine the previous picture, one can use a more precise procedure. Note that so far we focused on spatial locations of the nucleons in the precluster, ignoring the momentum distribution. That was possible because in a thermal state of nonrelativistic particles the kinetic and potential energy are simply additive, and momenta distributions are just Maxwell-Boltzmann's ones (with the corresponding effective mass for relative motion, $M_{\textrm{eff}}$).

The product of this Maxwell distribution and the spatial density matrix should be
projected to the \textit{ Wigner function} of the corresponding stationary states $W_0(x,p)$, the quantum analog of the phase space distribution,
\be \int d^3x \frac{d^3p}{(2\pi)^3} \ e^{-\frac{q^2}{2 M_{\textrm{eff}} T_f}} P(x,T) W_0(x,p) \ ,
\ee 
where $x$ is the relative coordinates and $q$ is relative momenta. Let us also note that for the temperature we should use the so-called kinetic freeze-out temperature $T_f \sim 100$ MeV. After the stage with $T_f$, there are---by definition---effectively no collisions, as witnessed by mesonic and baryonic $p_\perp$-spectra well explained by a convolution of hydrodynamic flow and thermal distributions~\cite{Adamczyk:2017iwn}.

This Wigner projection is not a new idea, and people using cascade or molecular dynamics codes for the description of heavy-ion collisions have been using it. However, this projection is customarily done by an oversimplified Gaussian form of the Wigner function~\cite{Nagle:1996vp}, 
\be W_0({\bf r},{\bf p})=8 \exp \left( -\frac{r^2}{d^2}-p^2 d^2\right) \label{Wigner_appr} \ , \ee
normalized to        
\begin{align}  1 & = \int d^3r \frac{d^3p}{(2\pi)^3} \ W_0({\bf r},{\bf p}) =\int d^3r \frac{d^3p}{(2\pi)^3} \ W_0^2({\bf r},{\bf p}) \ . 
\end{align}

The form~(\ref{Wigner_appr}) has only one parameter $d$, related to the r.m.s. radius. For the deuteron $d=1.7$ fm is usually used, corresponding to the r.m.s. deuteron radius of $2.1$ fm.
Furthermore, it was claimed that even dependence on the specific value of $d$ is rather weak, and that all what matters is that the  phase space volume has the right magnitude, corresponding to a single state. 

We call this approach ``oversimplified'' because it ignores the fact that
wave functions have at least two very different parts, ``in'' and ``out'' of the potential well.
Even the original approach to deuteron, by Bethe~\cite{Bethe}, via a rectangular attractive potential well, illuminated clearly  existence of   two distinct components of the wave function. An appropriate parametrization should have, at least, two Gaussians to be somewhat realistic.
The ``in'' component possesses large momenta related to the potential well depth $V$, the ``out'' has large size related to binding. Since $B\ll V$ they have different properties and do not correspond to the single Gaussian. Even larger difference should be present for multi-nucleon case. 

\begin{figure}[htp]
\begin{center}
\includegraphics[width=7cm]{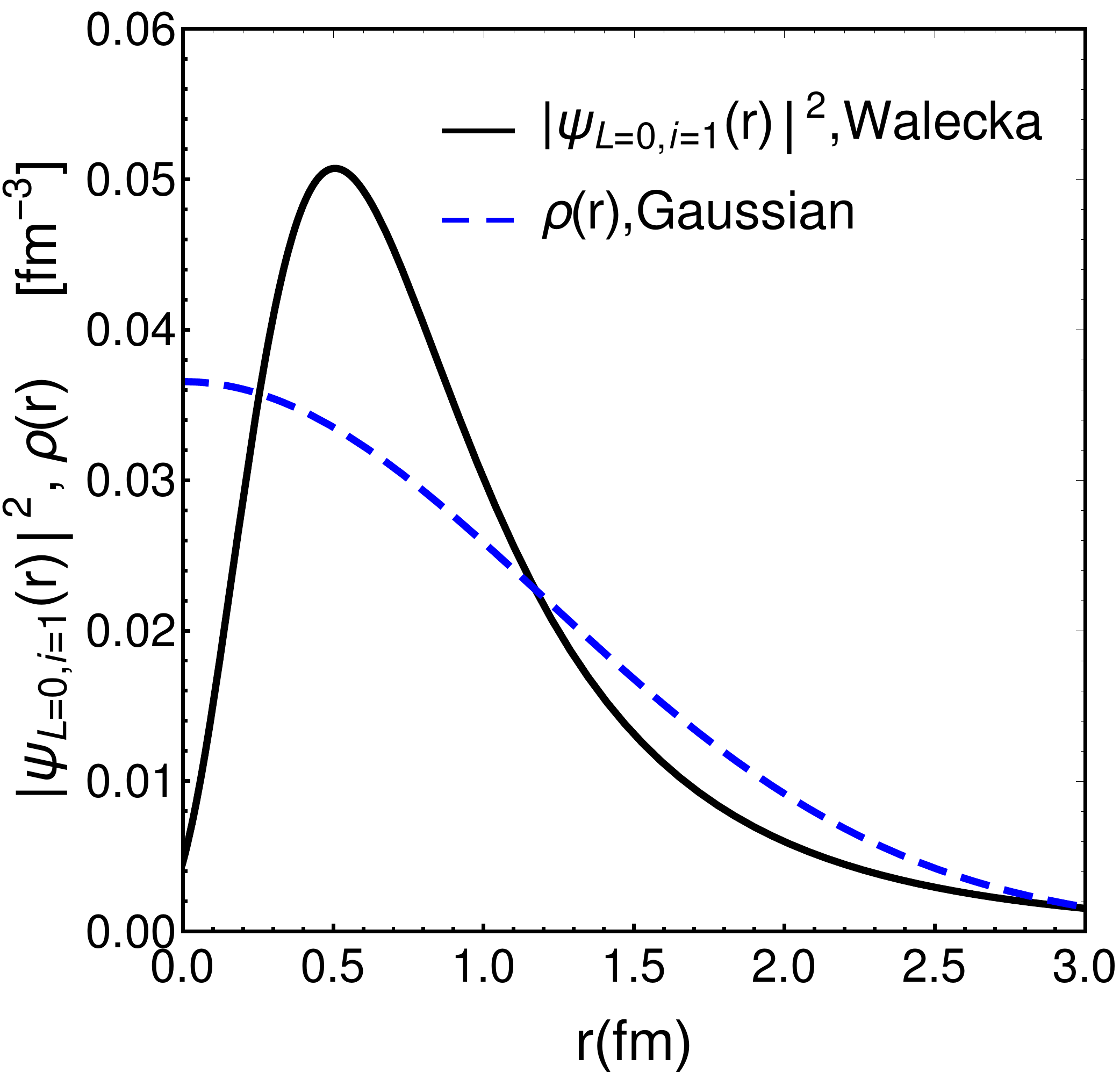}
% plot in WignerDeuteron.nb
\caption{Ground state wave function (squared) for the ``Walecka deuteron'' obtained numerically from Eq.~(\ref{eq:schrodinger}), and probability density Eq.~(\ref{eq:probden}) for the Gaussian Wigner function Eq.~(\ref{Wigner_appr}) of Ref.~\cite{Nagle:1996vp}.}
\label{fig:Wdeuteron}
\end{center}
\end{figure}

In Fig.~\ref{fig:Wdeuteron} we illustrate the ``Walecka deuteron'' wave function squared, $|\psi_{L=0,i=1} (r)|^2=|u_{L=0,i=1} (r)|^2/(4\pi r^2)$, obtained in Sec.~\ref{sec_2_nucl_Schr}, which is normalized as
\be \int d^3r |\psi_{L=0,i=1} (r)|^2 = 1 \ . \ee
We plot together the quantity~\cite{Case2008}.
\be \label{eq:probden} \rho({\bf r}) = \int \frac{d^3 p}{(2\pi)^3} \ W_0({\bf r},{\bf p})=\frac{\exp \left( -r^2/d^2 \right)}{\pi^{3/2}d^3} \ , \ee
using the Gaussian Wigner function in Eq.~(\ref{Wigner_appr}).
This probability density is, in fact, equal to the squared wave function~\cite{Case2008} of the deuteron. With the chosen normalization for the Wigner function, one has 
\be \int d^3r \rho({\bf r})=1 \ , \ee
so it makes sense to compare the square wave function obtained from Walecka potential and this probability density for a Gaussian wave function. We show the comparison in Fig.~\ref{fig:Wdeuteron}.

%Projecting our thermal preclusters, with Maxwell distribution of momenta of both nucleons and
%spatial correlations shown in Fig.\ref{fig_preclustering_pots}, 
% for the potentials we use, to  the Wigner function (\ref{Wigner_appr}) one finds that probabilities to form deuterons from them are 
% $ 0.16, 0.12,0.21,0.21$, for potentials $V_A,V_{B1}, V_{B2},V_{C}$, respectively.
%The complementary probabilities correspond to thermal precluster decaying into an
%unbound positive energy $pn$ pairs.
%
%(These probabilities should not be confused with coalescence parameters $B_A$ 
%used in description of spectra of deuterons and other light nuclei. Those parameters
%have dimension because they are inversely proportional to the volume of the fireball. The probabilities we calculated, we repeat for clarity, is not for a whole fireball but for a
%single thermal precluster.) 
%

\subsection{Possible observation of preclusters and statistical treatment of nuclear resonances~\label{sec:exictedhe4}}

Preclusters do not have fixed energy,  as they are superposition of physical states
in certain energy strip $\Delta E\sim T$. Being left alone, the preclusters decay into many physical states of the corresponding number of nucleons or light nuclei. In the previous section we focused on the precluster decay into the ground state. Now we discuss other decays (which of course dominate in terms of the total probability).

Let us consider as an example a  $ppnn$ precluster. Apart of forming a single bound state, the $\alpha$ particle or $^4$He, it can also decay into {\bf (i)} four individual nucleons; {\bf (ii)} 1+3 channels $p+ t,n+^3$He; {\bf (iii)} 2+2 channel $d+d$. The question then is whether one can experimentally infer the existence of preclusters by looking at these two-body channels.
 
One feature expected would be a peak at small relative momentum (rapidity). In the invariant mass distribution $(p_1+p_2)^2$ one also should find low-mass enhancement, related to feed-down from four-nucleon resonances. While we have not yet derived all of them from quantum mechanics, one can use those which were found experimentally. 
 
\begin{table}[ht]
\begin{center}
\begin{tabular}{|c|c|c|c|}
\hline
$E$ (MeV) & $J^P$ & $\Gamma$ (MeV)  & decay modes, in \% \\
\hline
\hline
20.21 & $0^+$ & 0.50  & $p$ = 100\\
21.01 & $0^-$ & 0.84  & $n$ = 24,  $p$ = 76\\
21.84 & $2^-$ & 2.01  & $n$ = 37,  $p$ = 63  \\
23.33 & $2^-$ & 5.01  & $n$ = 47,  $p$ = 53  \\
23.64 & $1^-$ & 6.20  & $n$ = 45,  $p$ = 55 \\
24.25 & $1^-$ & 6.10  & $n$ = 47,  $p$ = 50,  $d$ = 3 \\
25.28 & $0^-$ & 7.97  & $n$ = 48,  $p$ = 52\\
25.95 & $1^-$ & 12.66 & $n$ = 48,  $p$ = 52 \\
27.42 & $2^+$ & 8.69  & $n$ = 3,   $p$ = 3,   $d$ = 94 \\
28.31 & $1^+$ & 9.89  & $n$ = 47,  $p$ = 48,  $d$ = 5 \\
28.37 & $1^-$ & 3.92  & $n$ = 2,   $p$ = 2,   $d$ = 96 \\
28.39 & $2^-$ & 8.75  & $n$ = 0.2, $p$ = 0.2, $d$ = 99.6 \\
28.64 & $0^-$ & 4.89  & $d$ = 100 \\
28.67 & $2^+$ & 3.78  & $d$ = 100 \\
29.89 & $2^+$ & 9.72  & $n$ = 0.4, $p$ = 0.4, $d$ = 99.2 \\
\hline
\end{tabular}
\caption{Low-lying resonances of the $^4$He system, from BNL properties of nuclides.\footnote{
\url{https://www.nndc.bnl.gov/nudat2/getdataset.jsp?nucleus=4HE&unc=nds}} $J^P$ are the total angular momentum and parity,
$\Gamma$ is the decay width. The last column is the decay channel branching ratios, in percent. 
$p,n,d$ correspond to the emission of proton, neutron, or deuterons, respectively.
%  https://www.nndc.bnl.gov/nudat2/getdataset.jsp?nucleus=4HE&unc=nds
}\label{tab:he4}
\end{center}
\end{table}

In Table~\ref{tab:he4} we list such resonances occupying the strip of energies of width $\Delta E=10$ MeV above the binding threshold, shown with their quantum numbers and branching ratios for their decay modes.
 
Note that already in this strip the resonances are strongly overlapping, as the decay widths and energy differences are comparable. A growing density of states and widths above this strip
makes their separation/discovery hard. However, one does not need to find them one-by-one, 
but rather look for a collective enhancement near-zero effective mass .
 
In the spirit of the statistical thermal model, one may assume that all 
\be N_{\textrm{states}}=\sum_i (2 J_i +1) =49 \nn \ee 
states in this energy strip are populated $equally$ in the quantum decomposition 
of preclusters which in our classical simulation have corresponding energies.
With this assumption, and using the decays indicated in the table (interpreted as $p+t,n+^3$He, $d+d$
exclusive channels), one further finds that decays of a single $ppnn$ precluster should produce, on average, 0.30 ($p$ + tritium), 0.22\, ($n$ + $^3$He) and 0.96 deuterons (0.48 $dd$ pairs).  
Detector resolution permitting, one should search for evidences of these $p+t, d+d$
resonances in heavy-ion datasets. In particular, these evidences can only show up in the nuclear ratios we have been considering, as this ``feed-down'' is just a tiny effect in the absolute yields of nuclei. Should such ``feed down'' be found, it would obviously be a direct evidence for the four-nucleon preclustering we advocate in this work.

%%%%%%%%%%%%%%%%%%%%%%%%%%%%%%%%%%%%%%%%%%%%%%%%%%%%%%%%%%%%%%%%%%%%%%%%%%%%%%%%%%%%%%%%%%%%%%%%%%%
%%%%%%%%%%%%%%%%%%%%%%%%%%%%%%%%%%%%%%%%%%%%%%%%%%%%%%%%%%%%%%%%%%%%%%%%%%%%%%%%%%%%%%%%%%%%%%%%%%%
%%%%%%%%%%%%%%%%%%%%%%%%%%%%%%%%%%%%%%%%%%%%%%%%%%%%%%%%%%%%%%%%%%%%%%%%%%%%%%%%%%%%%%%%%%%%%%%%%%%

\section{Summary~\label{sec:summary}}

In our previous paper~\cite{Shuryak:2018lgd} we studied clustering of nucleons, at the freeze-out conditions of heavy-ion collisions, especially close to a possible critical point of QCD. The method used to simulate the real-time dynamics of nucleons, was a classical molecular dynamics code. Although for calculations in nuclear matter it was augmented by some phenomenological ``Fermi potential'' to mimic quantum effects, it was clear that a more quantitative study of few-body quantum mechanics was needed, as is indeed explored in the present paper. 

Before we come to their description, let us recall the main finding of Ref.~\cite{Shuryak:2018lgd}. It was shown that  the clustering phenomenon and its rate are extremely sensitive to even small modifications of the internucleon potential. The observable on which we focused in that paper was the scaled kurtosis of the (net-)proton multiplicity distribution, which was shown to be substantially increased by a reduction of the $\sigma$-mode mass.

Let us now come to the results of this paper, aiming first at experiment-oriented readers.  The
available  data on $N_t N_p/N_d^2$ ratio versus the collision energy, shown in Fig.~\ref{fig_ratio_tpdd}, are intriguing. At the highest RHIC energy this ratio is compatible with the ratio of statistical weights of a noninteracting gas (unit value on that plot). However, at lower energies it is about twice larger, perhaps with one (or two) maximum at certain collision energy. Since the main Boltzmann factors $\exp[(\mu_B-m_N)/T]$ cancel in the ratio, as well as thermal kinetic energy of four nucleons in numerator and denominator, any deviation from 1 should be assigned to some interaction. In particular, a stronger attraction in the three-nucleon system as compared to the two-nucleon one would bring this ratio to values larger than 1. An enhanced production of $t$ is thus interpreted above as a contribution from preclusters. 

If so, then we propose that similar effects, but enhanced, should be observed in other ratios including $^4$He like 
\be \frac{N_{\alpha} N_p}{N_{^3\textrm{He}} N_d} \ , \quad \frac{N_{\alpha} N_t N_p^2}{N_{^3\textrm{He}} N_d^3} \ , \quad \frac{N_\alpha N_p^2}{N_d^3} \ . \ee

The main object of this study, the four-nucleon preclusters, were found to be very interesting, even
for the unmodified $T =0$ nuclear forces. Out of $\sim 50$ bound states, only one---the ground state---is the observable $^4$He. All others have known decay channels as listed in Table~\ref{tab:he4}. We suggest that feed-down from them is also part of the reason for the enhanced $t$ production at low RHIC energies. One should study this suggestion experimentally, looking for explicit two-body decay channels of preclusters, as an enhancement at low invariant mass in, say $p+t, d+d$ channels. We also propose that the precluster decay into four protons is contributing to the enhanced kurtosis of the net-proton multiplicity distribution.

Now we turn to summary for readers interested in many-body theory. Among the goals of this paper are: 
\begin{enumerate}
\item [(i)] development of a novel semiclassical method for finite temperature density matrix, based on path integrals, called the thermal flucton;
\item [(ii)] comparing its results with classical Boltzmann distribution at high temperature, and with quantum ground state wave functions at low $T$; 
\item [(iii)] obtaining reliable estimates for precluster decay probabilities into $d$ for two nucleons, and $^4$He for four nucleons; 
\item [(iv)] obtaining estimates for two-body precluster decays, such as  $ppnn\rightarrow p+ t$, $n+^3$He, and $d+d$.
\end{enumerate}

We used first a (rather traditional) method to calculate the density matrix for four-nucleon system, via solving the Schr\"odinger equation for multiple energy levels, and weighting them by the Boltzmann factor. We did so for the two-nucleon system with a Serot-Walecka potential, and using the $K$-harmonics method for four nucleons. The results, shown in  Fig.~\ref{fig_kharmonics_density}, show a modest $\sim 1.4$ correlation for the unmodified potential, but $\sim 10$ enhancement for the modified one with increased attraction.

A part of this paper is devoted to the methodical development of the semiclassical ``flucton'' method~\cite{Shuryak:1987tr}, so far developed for $T=0$ only~\cite{Escobar-Ruiz:2016aqv,Escobar-Ruiz:2017uhx}. We have shown how to use it for nonzero temperatures. It does work well for standard toy models such as the anharmonic oscillator (see Fig.~\ref{fig_anharmonic}), and it is also applicable to two- and four-nucleon problem at finite temperatures. The flucton method (see Fig. \ref{fig_WalN4}) predicts somewhat larger effects than $K$-harmonics do, $\sim 4$ for the unmodified potential, and really huge enhancement for the modified one. The difference may be related to the fact that we only calculated the leading semiclassical part of the four-nucleon density matrix, $\exp(- S_{\textrm{flucton}})$, without the one-loop pre-exponent (determinant) or other corrections. It may also indicate that the action is not large enough to fully trust the semi classical approach. 

\vspace{4mm}

{\it Note added in proof:}  The large kurtosis at the lowest energy central Au+Au collisions observed by STAR collaboration (which triggered our study of four-nucleon systems) was recently also observed by the HADES collaboration~\cite{Adamczewski-Musch:2020slf} at even lower energy $\sqrt{s_{NN}}=2.4$ GeV. The feed-down from four-nucleon resonances, that we suggested in this paper, has been applied in Ref.~\cite{Lorenz} using a new statistical-thermal model, improving the description of HADES hadron production data.

%%%%%%%%%%%%%%%%%%%%%%%%%%%%%%%%%%%%%%%%%%%%%%%%%%%%%%%%%%%%%%%%%%%%%%%%%%%%%%%%%%%%%%%%%%%%%%%%%%%
%%%%%%%%%%%%%%%%%%%%%%%%%%%%%%%%%%%%%%%%%%%%%%%%%%%%%%%%%%%%%%%%%%%%%%%%%%%%%%%%%%%%%%%%%%%%%%%%%%%
%%%%%%%%%%%%%%%%%%%%%%%%%%%%%%%%%%%%%%%%%%%%%%%%%%%%%%%%%%%%%%%%%%%%%%%%%%%%%%%%%%%%%%%%%%%%%%%%%%%

\begin{acknowledgments}

This work was supported in part by the Office of Science, U.S. Department of Energy under Contract No. DE-FG-88ER40388. J.M.T.-R. also acknowledges financial support from the Deutsche Forschungsgemeinschaft (DFG, German Research Foundation) through Projects No. 411563442 (Hot
Heavy Mesons) and No. 315477589 - TRR 211 (Strong-interaction matter
under extreme conditions).

\end{acknowledgments}

\appendix

\section{Wave function of $^4$He using $K$-harmonics} \label{app_K_harmonics}

The so-called method of $K$-harmonics was developed in Ref.~\cite{Badalian:1966wm}. Its main idea is that the multi-dimensional Schr\"odinger equation can be treated with a single ``radial'' coordinate plus ``angular variables'', for which a complete set of functions is known. In certain cases a rather good approximation can be obtained using a single lowest angular function, with trivial angular dependence. Such cases include in particular $A=3$
nuclei and also  $^4$He, which is the case we will discuss here following Ref.~\cite{KHarmonics_he4}. Since these papers are rather old, we indicate in this Appendix their main points.

As a preliminary information, let us note that $^4$He is a surprisingly compact nucleus, with a r.m.s. radius of only $R(^4\textrm{He})\approx 1.6$ fm. Its binding energy may appear
to be large $B(^4\textrm{He})=28.3$ MeV, but since there are six nucleon pairs the ``binding per pair'' is rather small and only about twice that of the deuteron. 

The first standard step in many-body physics is the separation of the center of mass motion from the relative coordinates. It is usually done using Jacobi coordinates, which for the $A=4$ case are
\begin{align} 
 \vec \xi_1 & =\frac{\vec x_1 - \vec x_2}{\sqrt{2}}, \,\,\, \vec \xi_2=
\frac{\vec x_1 + \vec x_2 - 2 \vec x_3}{\sqrt{6}} \ , \\ 
\vec \xi_3 & = \frac{ \vec x_1 + \vec x_2 +\vec x_3 - 3 \vec x_4}{2 \sqrt{3}} \ . 
\end{align}
The radial coordinate, or hyperdistance, is defined as
\be \rho^2=\sum_{m=1}^3  (\vec \xi_m)^2=\frac{1}{4} \left[ \sum_{i\neq j} (  \vec x_i-\vec x_j)^2 \right] \ .
\label{rho_definition}
\ee
The radial part of the Laplacian in these Jacobi coordinates is $\psi''(\rho)+8\psi'(\rho)/\rho$,
and using the substitution 
\be \chi(\rho)=\psi(\rho) \rho^4 \label{substitution} \ , \ee
one arrives to the conventional-looking Schr\"odinger equation for $K=0$ harmonics
\be 
\frac{d^2 \chi}{d\rho^2} -\frac{12}{\rho^2} \chi-\frac{2m_N}{\hbar^2} [W(\rho)+V_C(\rho)-E] \chi=0 \ ,
\label{eqn_radial_for4}
\ee
where $W$ is the projection of the potential to this harmonic. According to Ref.~\cite{KHarmonics_he4},
\be W(\rho)=\frac{315}{4} \int_0^1 V_{NN}(\sqrt{2}\rho x)  (1-x^2)^2 x^2 dx \ , \ee
where $V_{NN}(r)$ is the two-body nuclear potential.

We consider the simplest nuclear potential of Ref.~\cite{KHarmonics_he4} (called $V_1$ there) 
\be \label{eq:VNNK} V_{NN}(r)=-83.34 \ e^{- r^2/1.6^2}+144.86 \ e^{-r^2/0.82^2} \ , \ee
with the prefactors given in MeV while the radii in the exponents in fm.
In Eq.~(\ref{eqn_radial_for4}) also appears a Coulomb repulsion between the two protons, which adds $V_{C} (\rho)=2.23$ MeV$\cdot$ fm$/\rho$.
The discussion of the solutions of this equation is given in the main text, where not only the ground state but also the first $J^P=0^+$ excitation can be identified with physical states.

\begin{figure}[htp]
\begin{center}
\includegraphics[width=7cm]{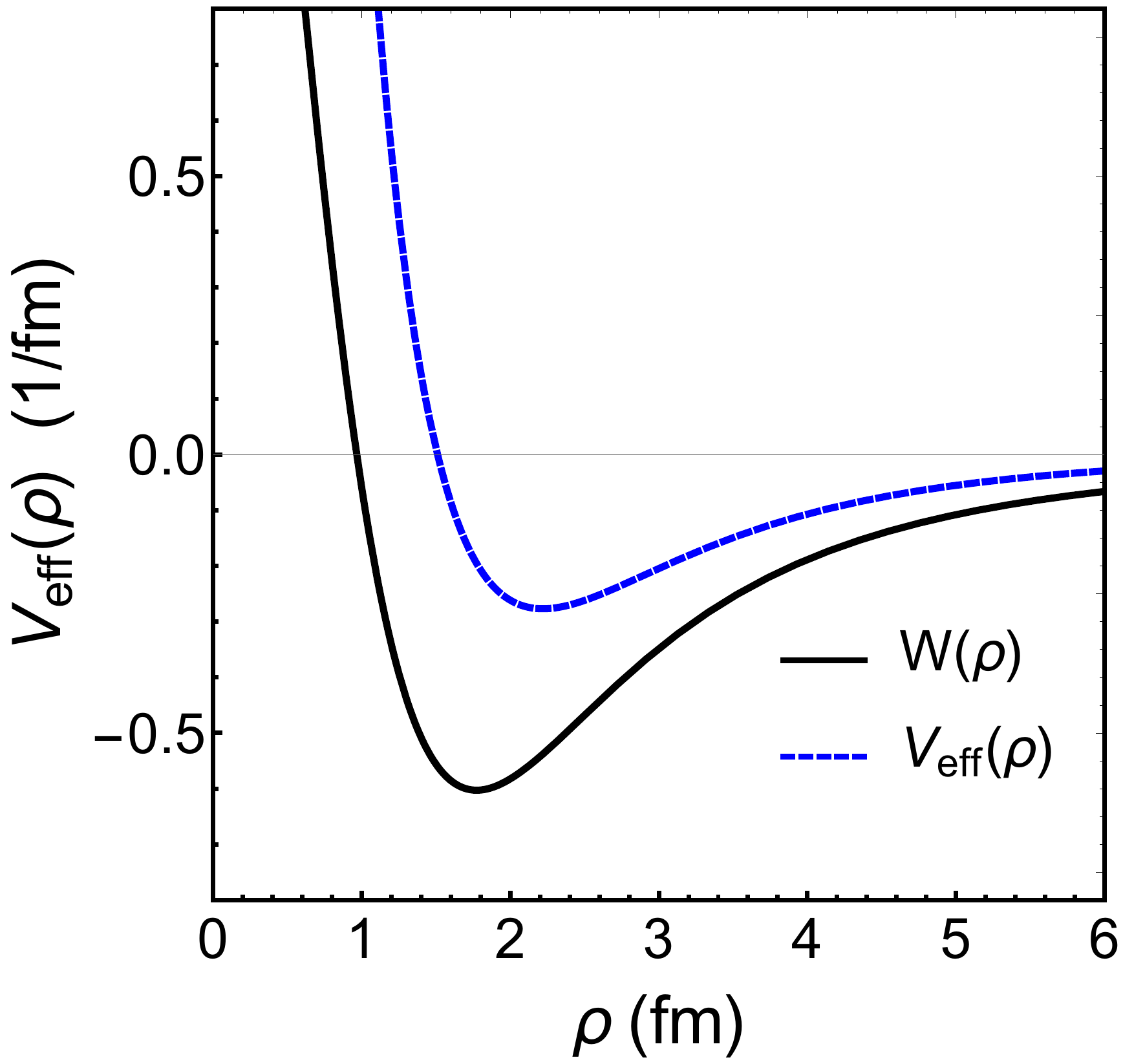}
\caption{$W(\rho)$ and $V_{\textrm{eff}} (\rho)$ potentials used in Ref.~\cite{KHarmonics_he4} for the solution of Eq.~(\ref{eqn_radial_for4}). The second potential is also used in the semiclassical solution in Sec.~\ref{sec:fluchyper} using the flucton path. }
\label{fig:Wpot}
\end{center}
\end{figure}

For the application of the problem Eq.~(\ref{eqn_radial_for4}) into the semiclassical flucton solution, it is easy to realize that it is equivalent to a 1D Schr\"odinger equation,
\be - \frac{\hbar^2}{2m_N} \frac{d^2 \chi}{d\rho^2} + V_{\textrm{eff}}(\rho) \chi = E\chi \ , \ee
with the effective potential,
\be \label{eq:Veff} V_{\textrm{eff}} (\rho)=W( \rho) +\frac{6\hbar}{m_N\rho^2} +V_{C} (\rho)  \ . \ee
Therefore we can apply the standard flucton method described in the text to obtain the flucton solution to the inverted potential $-V_{\textrm{eff}} (\rho)$.
The potentials $W(\rho)$ and $V_{\textrm{eff}} (\rho)$, for the special case of the $NN$ potential in Eq.~(\ref{eq:VNNK}), are plotted in Fig.~\ref{fig:Wpot}.

\section{Semiclassical theory at finite temperature}  \label{app_fluctons}

In this Appendix we illustrate how the flucton method is applied for the 1D harmonic oscillator problem, with the Euclidean action
\be S_E [x(\tau)]=\int d\tau \left( \frac{\dot x^2}{2} + \frac{x^2}{2} \right) \ , \ee
where three mechanical units are chosen to have $\hbar=m=\omega=1$. The dot indicates derivative over the Euclidean time $\tau=i t$, and the circle at the integral reminds us that it is defined on a Matsubara circle.  
Note that the sign of the potential in the action is reversed, which is the consequence of $i^2=-1$ in the kinetic term. 

The flucton is a classical path which: {\bf (i)} passes through some observational point $x_0$; 
and {\bf (ii)} is periodic with the period $\beta$ in $\tau$. At zero temperature, because in Euclidean time the potential is inverted, the particle is ``sliding'' from the maximum at $\tau=0$ to $\tau=\pm \infty$. Most of the previous applications were at $T=0$ ($\beta=\infty$) and the slide was always started from the maximum, at zero energy.

At nonzero $T$ such slides also start with zero velocity but from a certain ``turning point'' $x_{\textrm{turn}}$ and proceed toward the observational point $x_0$. The turning point, by symmetry, should be separated from $x_0$ by the time equal to half period $\beta/2$. For any one-dimensional motion there is no need to use the Newton's equation of motion. Expressing the velocity from the energy conservation on the path, this condition can be put into the general form
\be \frac{\beta}{2}= \int_{x_{\textrm{turn}}}^{x_0} \frac{dx}{\sqrt{2(V(x)+E)/m}} \label{eqn_period} \ . \ee
For the harmonic oscillator, with $V(x)=x^2/2$, it is easy to find the turning point by solving
\be E=V(x_{\textrm{turn}})=\frac{x_{\textrm{turn}}^2}{2} \ , \ee
and calculate the integral for the period
\be  \frac{\beta}{2}=\textrm{arccosh} \left( \frac{x_0}{\sqrt{2E}}\right) \ . \ee

The classical flucton path is therefore given by
\be x_{\textrm{fluc}}(\tau)=x_0 \frac{ \textrm{cosh} (\tau-\beta/2)}{ \textrm{cosh} (\beta/2)} \ , \ee
and at both $\tau=0$ and $\tau=\beta$ it returns to the desired point $x_0$. 
Now, substituting this solution into the Euclidean action one finds that 
\be S_E [x_{\textrm{fluc}}(\tau)]=x_0^2 \ \textrm{tanh} \left( \frac{\beta}{2} \right) \ , \ee
and the density matrix is therefore Gaussian at all temperatures
\be P(x_0) \sim e^{-S_E[x_{\textrm{fluc}}(\tau)]} = e^{-x_0^2 \ \textrm{tanh} \left( \frac{\beta}{2} \right) } \ . \ee
This reproduces the result obtained by Feynman~\cite{Feynman_SM} via the explicit calculation of the Gaussian path integral. As it happens for any Gaussian path integral, this semiclassical formula is, in fact, exact.

Let us now proceed to illustrate the first nontrivial problem, the anharmonic oscillator, defined by 
\be S_E [x(\tau)]=\int d\tau \left( \frac{\dot x^2}{2} + \frac{x^2}{2} + \frac{g}{2} x^4 \right) \ . \ee 
The tactics used in the previous example are not easy to implement: in particular,
the period condition Eq.~(\ref{eqn_period}) defining the energy $E$ needs to be solved numerically
for each value of the $x_0$. Furthermore, using energy conservation leads naturally to
$\tau(x)$ representation of the path, rather than the conventional $x(\tau)$.

After trying several strategies we concluded that the simplest way to solve the problem is: 
\begin{itemize}
 \item [(i)] solve numerically the second-order equation of motion, 
 \be \ddot{x} = \frac{\partial V(x)}{\partial x} = x + 2gx^3 \ , \ee
 starting not from the 
observation point $x_0$ but from the turning point $x_{\textrm{turn}}$ at $\tau=-\beta/2$. This is easier because the velocity vanishes at this point, and a numerical solver can readily be used;
 \item [(ii)] follow the solution for half period  $\beta/2$ and thus
find the location of  $x_0=x(\tau=0)$;
 \item [(iii)] calculate the corresponding action and double it, to account for the other half period $\tau\in (0,\beta/2)$.
\end{itemize}

Notice that this method provides $x_0$ as an \textit{ output} after solving the equations of motion with initial conditions $x(-\beta/2)=x_{\textrm{turn}}$ and $\dot{x}(-\beta/2)=0$. One could also tweak a bit the method to use $x_0$ it as an \textit{ input} by using the constraints $x(0)=x_{0}$ and $\dot{x}(-\beta/2)=0$. The details of this procedure and its comparison with the numerical results based on the definition Eq.~(\ref{eqn_P_standard}) for the anharmonic oscillator will be provided in a separate methodical paper~\cite{T_fluctons}.

\begin{figure}[htp]
\begin{center}
\includegraphics[width=7cm]{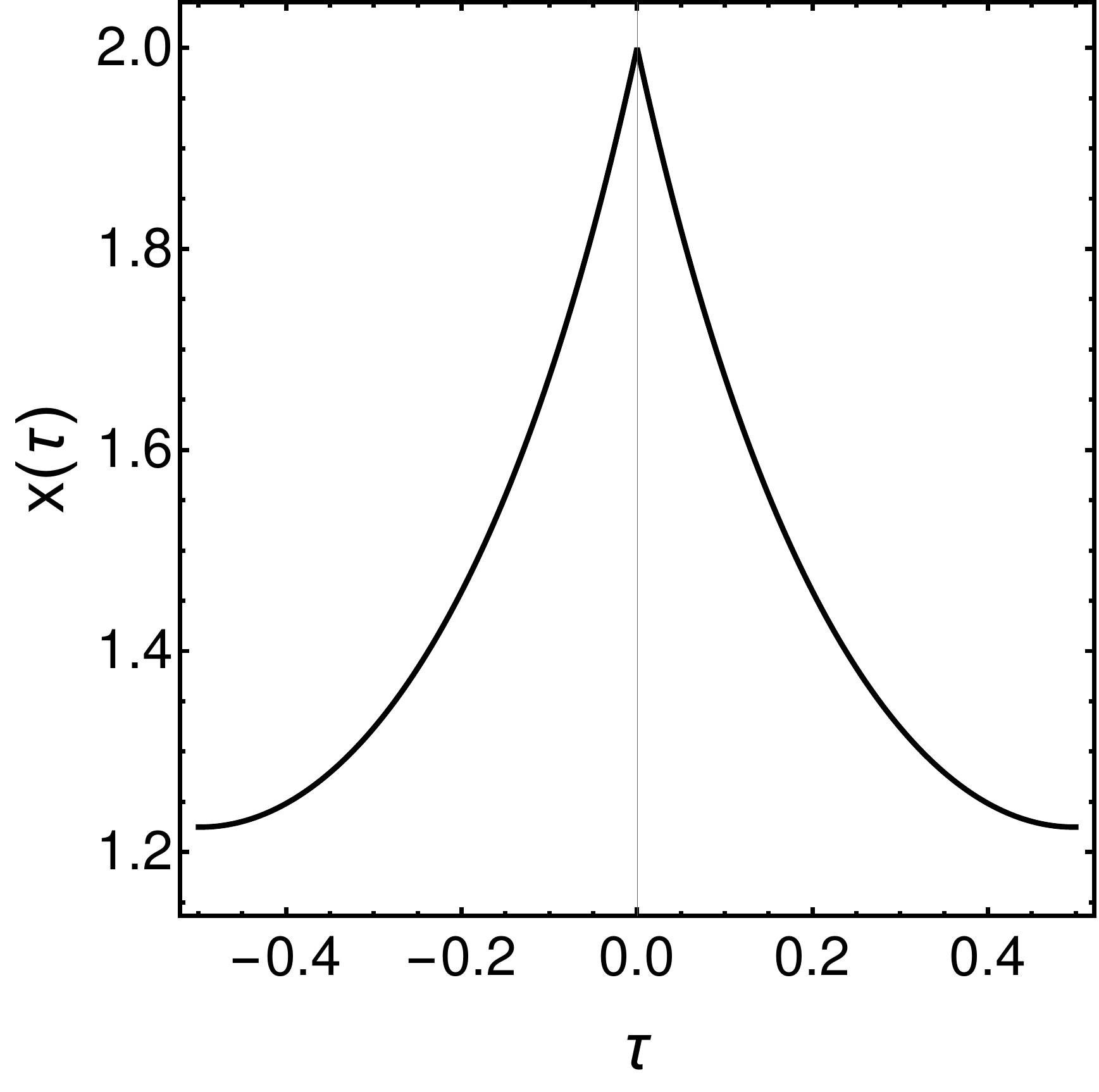}
\caption{Flucton path for the anharmonic oscillator with $g=1$ and $T=1$ (in units of the mass), for the observation point $x_0=2$. Notice that, as expected, $\tau \in (-\beta/2,\beta/2)$ with $\beta=1/T=1$ and $x(\tau=0)=x_0$.}
\label{fig:flucAHO}
\end{center}
\end{figure}

In Fig.~\ref{fig:flucAHO} we show the numerical solution of the flucton path for the anharmonic oscillator with $g=1$ and $T=1$ (in units of the mass). We choose the observation point $x_0=2$, which is reached as expected, at $\tau=0$ (cf. Fig.~\ref{fig_fl}). The flucton is periodic in $\tau$ with period $\beta=1/T$.

\begin{figure}[htp]
\begin{center}
\includegraphics[width=7cm]{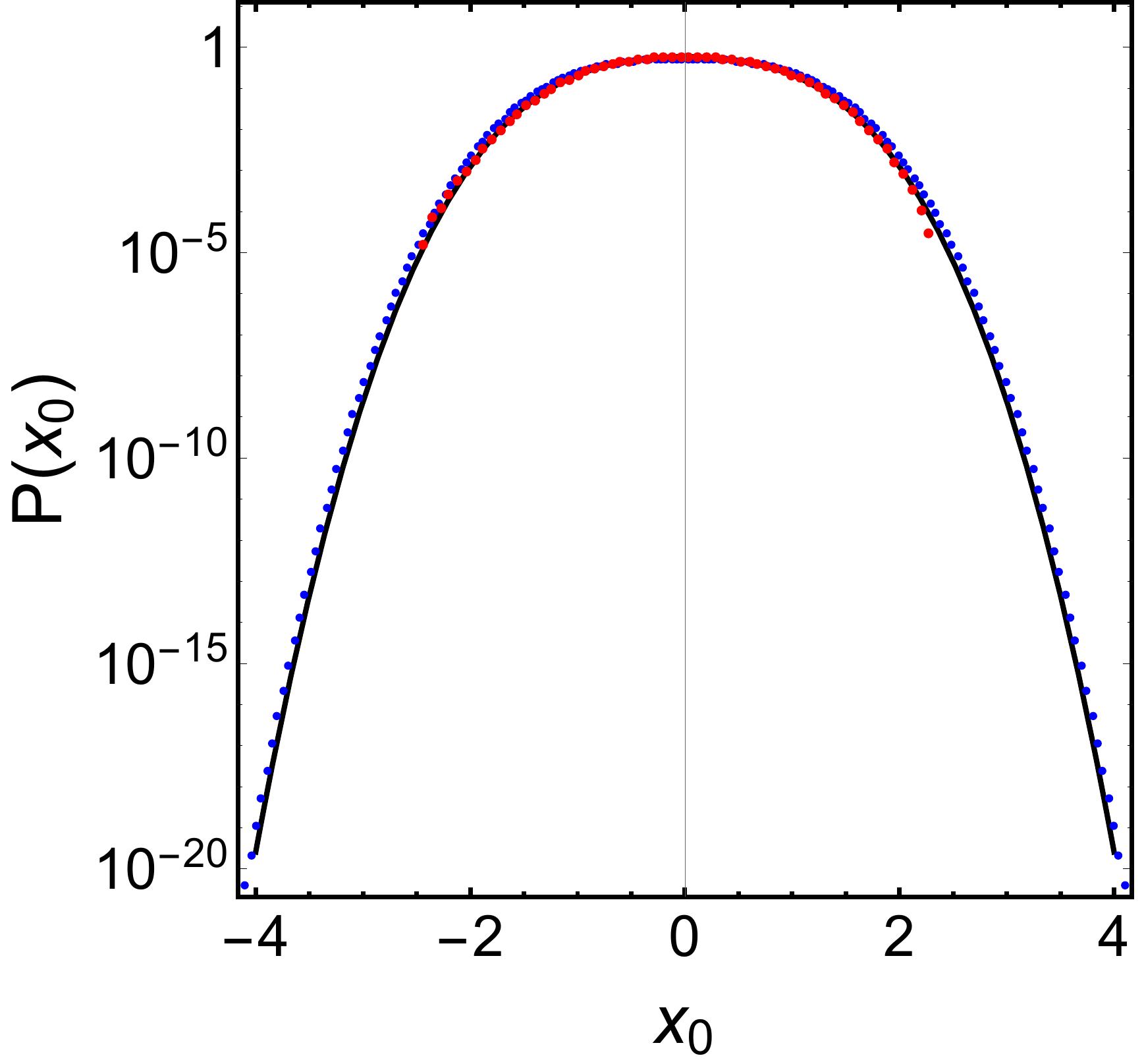}
\includegraphics[width=7cm]{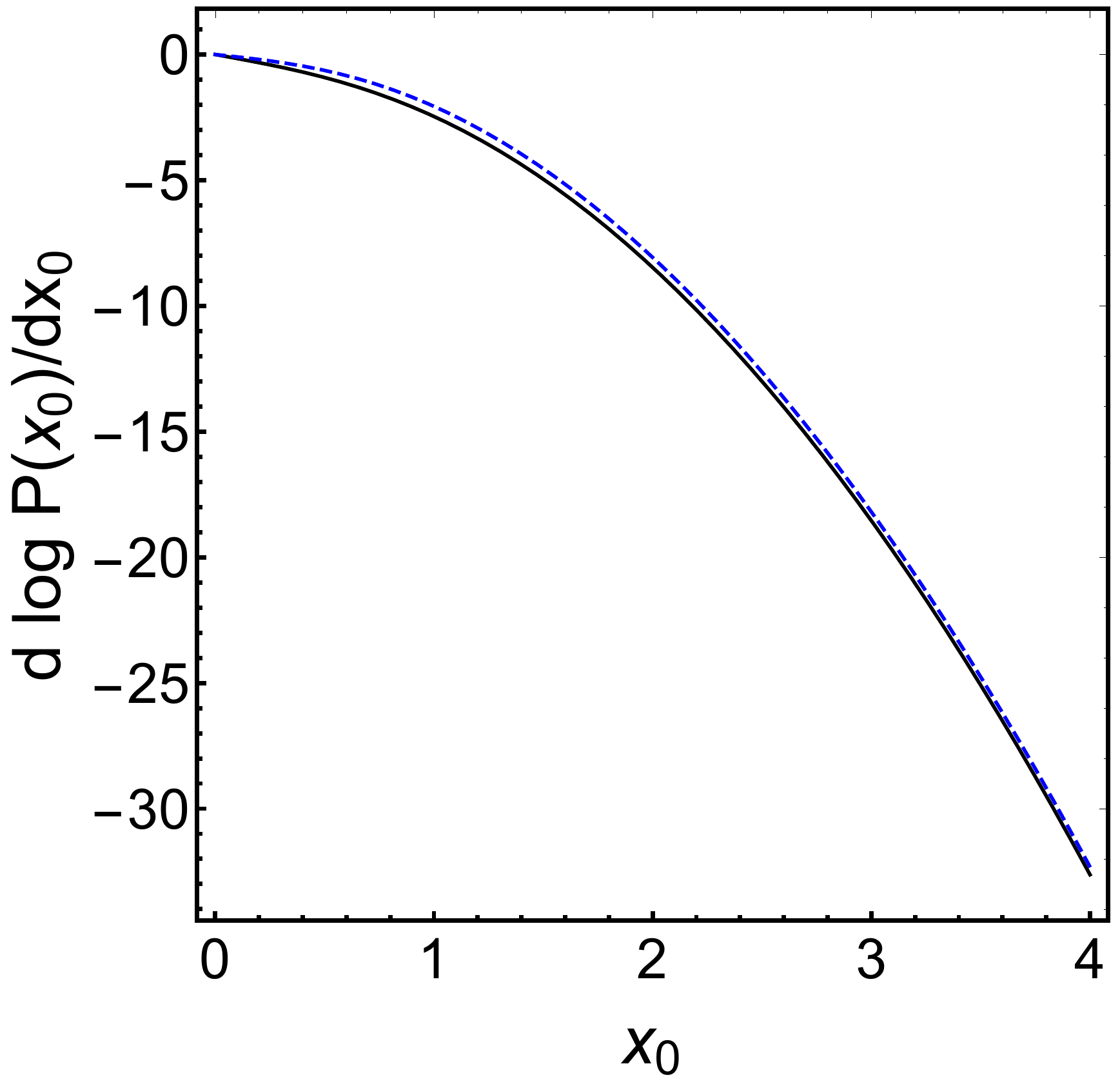}
\caption{Top panel: Density matrix $P(x_0)$ vs $x_0$ for anharmonic oscillator with the coupling  $g=1$,
at temperature $T=1$, calculated via the definition Eq.~(\ref{eqn_P_standard}) (line) and
the flucton method (points). The line is based on 60 lowest state wave functions found
numerically. Bottom panel: Comparison of the logarithmic derivative of the density matrix of the upper panel.}
\label{fig_anharmonic}
\end{center}
\end{figure}

Here we present the upper panel of Fig.~\ref{fig_anharmonic} comparing the summation over 60 squared wave functions, and Boltzmann weighted (solid line),  with the result of the flucton method (points) at $T=1$ (in units of the mass). The coupling is set to $g=1$. For additional comparison we also present the numerical results of a path integral Monte Carlo calculation with the same parameters which simulates quantum paths of one particle in the anharmonic oscillator potential. The method is inspired by the nice reference Ref.~\cite{Ceperley:1995zz} and will be reviewed in Ref.~\cite{T_fluctons}.

As a semiclassical approach one expects that the flucton solution works better when the action is large, i.e. for large values of $x_0$. However, one observes that the flucton systematically overestimates the solution based on the Schr\"odinger solution. Part of the discrepancy comes from normalization issues as described in~\cite{Escobar-Ruiz:2017uhx}. To remove those it is enough to compare the logarithmic derivative of the density matrix $d \log P(x_0)/dx_0$. In the bottom panel of Fig.~\ref{fig_anharmonic} we show the logarithmic derivative of the density matrix in linear scale. While the agreement is nearly perfect, a small difference can still be detected. We ascribe it to the loop corrections of the thermal flucton solution~\cite{Escobar-Ruiz:2017uhx}.

\end{document}